\documentclass[%
reprint,
superscriptaddress,
nofootinbib,
amsmath,amssymb,
aps,
prd,
]{revtex4-2}

\usepackage{graphicx}
\usepackage{dcolumn}
\usepackage{bm}
\usepackage[mathlines]{lineno}
\usepackage[normalem]{ulem}
\usepackage{cancel}
\usepackage{upgreek}
\usepackage{lipsum}
\usepackage{tensor}
\usepackage{mathtools}
\usepackage[utf8]{inputenc}
\usepackage{orcidlink}
\usepackage{hyperref}
\hypersetup{
    colorlinks = true,
    citecolor = blue,
    urlcolor=blue, linkcolor=blue}
\usepackage[capitalise]{cleveref}

\newcommand{\order}{\mathcal{O}}
\usepackage{letltxmacro}
\newcommand{\munu}{{\mu \nu}}

\crefname{section}{Sec.}{Sec.}

\begin{document}

\preprint{APS/123-QED}

\title{Beyond black hole spectroscopy: Quasinormal mode contamination by massless scalars
}

\author{Miguel Yulo Asuncion\,\orcidlink{0009-0003-5075-3107}}
\affiliation{Nottingham Centre of Gravity \& School of Mathematical Sciences, Nottingham NG7 2RD, United Kingdom} 
\author{Giovanni D'Addario\,\orcidlink{0009-0006-5112-2595}}
\affiliation{Nottingham Centre of Gravity \& School of Mathematical Sciences, Nottingham NG7 2RD, United Kingdom} 
\affiliation{School of Physics and Astronomy, University of Nottingham, Nottingham NG7 2RD, United Kingdom}
\author{Thomas P. Sotiriou\,\orcidlink{0000-0002-9089-4866}}
\affiliation{Nottingham Centre of Gravity \& School of Mathematical Sciences, Nottingham NG7 2RD, United Kingdom} 
\affiliation{School of Physics and Astronomy, University of Nottingham, Nottingham NG7 2RD, United Kingdom}

\date{\today}

\begin{abstract}

Testing General Relativity (GR) with black hole ringdowns has conventionally focused on attempting to detect shifts away from the quasinormal mode (QNM) frequencies of the Kerr metric. It has recently been argued, however, that the ringdown signal will also be contaminated with the QNM frequencies of any new fields that are present in a beyond-GR scenario, provided that they couple nonminimally to gravity. We study black hole perturbations for the shift-symmetric Horndeski action, which includes all interactions between a massless scalar and gravity that lead to second order equations upon variation. We perturb linearly in the field and also employ a perturbative expansion in the scalar charge per unit black hole mass, $q$. Assuming that the scalar amplitude is suppressed by $q$, we demonstrate that, to order $q^2$, the coupling between the scalar and the Gauss-Bonnet invariant is the only term that contributes to both frequency shifts and contamination, and that the two effects appear at the same perturbative order. If the assumption about the suppression of the scalar amplitude is relaxed, contamination can appear at leading order in $q$, and hence dominate over frequency shifts. In this case, contamination also receives subleading corrections from an additional coupling constant.

\end{abstract}

\maketitle

\section{\label{sec:Introduction}Introduction}

Building upon over a century of experimental verifications of General Relativity (GR) \cite{RevModPhys.19.361, Gilmore_2021, PhysRevLett.4.337, 1972Sci...177..166H, 1975ApJ...195L..51H, PhysRevLett.106.221101}, the direct detections of gravitational waves (GWs) by the LIGO-Virgo-KAGRA (LVK) collaboration \cite{Abbott_2016, Abbott_2021, Abac_2025} has inaugurated a new regime for exploring the astrophysical consequences of Einstein's theory. The now routine detections of GW signals arising from the \textit{inspiral}, \textit{merger}, and post-merger \textit{ringdown} phases of binary black hole coalescences continue to provide ever more stringent tests of GR  \cite{PhysRevLett.116.221101, Abbott_2019, 6c61-fm1n}. The endeavors of the network of gravitational wave detectors will be bolstered by planned next-generation detectors such as the Einstein Telescope \cite{Punturo_2010, Abac_2026}, Cosmic Explorer \cite{2019BAAS...51g..35R, Evans:2021gyd}, and the spaceborne LISA detector \cite{amaroseoane2017laserinterferometerspaceantenna}.  

Within GR, the \textit{no-hair} theorems state that a black hole is exclusively described by its mass, spin, and electric charge \cite{PhysRev.164.1776, PhysRevLett.26.331, PhysRevLett.34.905}. Astrophysically, however, electric charge is assumed to be negligible, and thus black holes in nature are described by the Kerr metric \cite{PhysRevLett.11.237}.
 A consequence of this fact is that mass and spin uniquely characterize the quasinormal mode (QNM) frequencies $\omega_{\text{GR}}$ of ringdown signals from such Kerr black holes \cite{10.1098/rspa.1985.0119, Hans-Peter_Nollert_1999, Berti_2009}. 
The dynamical metric perturbations describing the QNM signal in this post-merger phase can be expressed at a fixed point as a sum of modes
\begin{equation}\label{eq:GR QNM}
    h_{\mu \nu} \sim \sum_{\mathbf{n}} A_\mathbf{n}  \mathrm{e}^{-\mathrm{i}  \omega_{\text{GR},\mathbf{n}}t},
\end{equation}
where the shorthand notation $\mathbf{n}$ denotes the angular momentum number $\ell$, the azimuthal number $m$, and the overtone number $n$. Each mode is characterized by an amplitude $A_\mathbf{n}$ and a frequency $ \omega_{\text{GR},\mathbf{n}}$.

While GR has passed numerous tests, issues such as the dark matter problem at galactic scales \cite{Corbelli_2000, Rogstad1972}, the dark energy problem at cosmological scales \cite{RevModPhys.61.1, 1998AJ....116.1009R}, and the need for a UV completion of gravity at quantum scales strongly suggest that there is gravitational physics that lies outside the realm of GR. This motivates the study of theories beyond GR (bGR) (e.g.~\cite{PhysRev.124.925, RevModPhys.82.451,Horndeski:1974wa,  Clifton:2011jh}), which can be used as effective field theories that capture part of this new physics. Such theories typically include new fundamental fields that modify the geometry of black hole spacetimes, in which case the black holes are said to possess \textit{hair}.

As the QNMs from the ringdown of such hairy black holes would thus also be modified by the new fields, analysis of ringdown signals may reveal signatures of hair and thus evidence of bGR theories \cite{PhysRevD.81.124021, PhysRevD.94.104024}. Typically, within the black hole spectroscopy program \cite{Dreyer:2003bv,berti2006gravitational}, these deviations from the GR ringdown are modelled by a \textit{frequency shift}: the frequency of each mode changes as $\omega_{\text{bGR},\mathbf{n}} = \omega_{\text{GR},\mathbf{n}} + \delta \omega_{\mathbf{n}}$, such that the metric perturbations are now expressed as
\begin{equation}\label{eq:Shift QNM}
   h_{\mu \nu} \sim \sum_\mathbf{n} A_\mathbf{n} \mathrm{e}^{-\mathrm{i}  (\omega_{\text{GR},\mathbf{n}} + \delta \omega_{\mathbf{n}})t}.
\end{equation}
It has been argued, however, that in addition  to the shift of the GR metric perturbation's QNM frequencies, the dynamical modes of any extra fields would themselves contribute to the ringdown signal with their own frequencies $\hat\omega_\mathbf{n}$ \cite{vdk4-tg61, SciPostPhys.20.1.025, ptg5-f769}. This  appears as a new contribution to  the metric perturbation,
\begin{equation}\label{eq:Contamination QNM}
    h_{\mu \nu} \sim \sum_\mathbf{n} \Big[ A_\mathbf{n}  \mathrm{e}^{-\mathrm{i}  (\omega_{\text{GR},\mathbf{n}} + \delta \omega_\mathbf{n})t} + B_\mathbf{n} \mathrm{e}^{-\mathrm{i}  \hat{\omega}_\mathbf{n}t}\Big] , 
\end{equation}
an effect referred to as \textit{frequency contamination} \cite{vdk4-tg61}. Here, $B_\mathbf{n}$ are the independent amplitudes of the contaminating modes, and $\hat{\omega}_\mathbf{n}$ their frequencies.

As tests for bGR corrections and parameter estimations from LVK ringdown data have been based on models that only include the frequency shift effect \cite{ 6c61-fm1n, kw5g-d732, LIGOScientific:2026wpt, PhysRevD.107.044030, 2025arXiv250614695K}, current methods of analysis may possess a systematic bias from missing the effects of contamination \cite{vdk4-tg61, SciPostPhys.20.1.025, ptg5-f769}.
Using the modified Teukolsky equation \cite{Li:2022pcy,Hussain:2022ins} (an extension of the seminal work of Teukolsky \cite{1973ApJ...185..635T} in GR to a broad class of bGR theories), Ref.~\cite{vdk4-tg61} determined the general, theory-agnostic ansatz for the modelling of the ringdown with QNMs, including both frequency shifts and contamination. 
Strikingly,
the contamination actually appears at the same order as the frequency shift in a perturbative expansion in the dimensionless coupling constant that controls the deviations from GR. 

It is instructive to apply such an analysis to a specific theory that allows for hairy black holes, in order to elucidate how both frequency shifts and contamination come about and which interactions of the new field control the magnitude of each effect. Here we will focus on shift-symmetric Horndeski gravity \cite{Horndeski:1974wa, Deffayet_2013,PhysRevD.90.124063,PhysRevD.99.124004} (invariant under transformations $\phi \rightarrow \phi  + \text{constant}$). This is the most general action that leads to second-order field equations for a massless scalar field $\phi$ that is nonminimally coupled to gravity. This choice is partly motivated by the fact that ultralight scalars are ubiquitous in extensions of GR \cite{Clifton:2011jh,Sotiriou:2014yhm,Barack:2018yly} and the Standard Model \cite{Copeland:2006wr,Kim:2008hd,Marsh:2015xka,Hui:2021tkt}, and partly because the massless limit is what was studied in earlier work  \cite{PhysRevD.109.084046,vdk4-tg61}. 

Scalar fields are covered by no-hair theorems \cite{Hawking:1972qk,PhysRevLett.108.081103, doi:10.1142/S0218271815420146, Hui:2012qt}. For shift-symmetric scalars, an exception arises when the scalar field is linearly coupled to the Gauss-Bonnet invariant 
\begin{equation}\label{eq:Gauss-Bonnet}
    \mathcal{G}=R^{\mu \nu \lambda \kappa} R_{\mu \nu \lambda \kappa}-4 R^{\mu \nu} R_{\mu \nu}+R^2,
\end{equation}
where $R_{\mu \nu \lambda \kappa}$ is the Riemann tensor, $R_{\mu \nu}$ the Ricci tensor, and $R$ the Ricci scalar \cite{PhysRevLett.112.251102, PhysRevD.90.124063}. Other interactions will have a subdominant effect on the scalar configuration \cite{Thaalba:2022bnt} and on the scalar charge per unit mass $q$ \cite{PhysRevD.99.124004}. It has been shown that the latter can be expressed as \cite{PhysRevD.99.124004}   
\begin{equation}\label{eq:scalar charge}
 q=\frac{1}{4 \uppi}\frac{\alpha}{M^2} \int_{\mathcal{H}} n_\mu \mathcal{G}^\mu,
\end{equation}
where $M$ is the mass of the black hole, $\alpha$ is the coupling constant governing the strength of the $\phi \mathcal{G}$ interaction, $n_\mu$ is the normal to the horizon $\mathcal{H}$, and $\mathcal{G}^\mu$ may be defined from \cref{eq:Gauss-Bonnet} as $\mathcal{G} \equiv \nabla_\mu \mathcal{G}^\mu$. The contributions to the integral by interactions other than the linear coupling with ${\cal G}$ are suppressed by the corresponding (dimensionful) coupling. Hence, one can infer that, to leading order in inverse powers of $M$, the magnitude of $q$ is inversely proportional $M^{2}$.

In this paper, we employ the formalism developed in \cite{PhysRevD.109.084046} to study how the frequency shift and contamination effects arise in shift-symmetric Horndeski gravity. The formalism involves a double expansion, with linear order dynamical perturbations controlled by $\epsilon$, as well as stationary corrections in $q$, to both the metric and the scalar field. Given the scaling of $q$ with $M$, this expansion  can also be seen as a large black hole mass expansion. We now extend the formalism to ${\cal O}(\epsilon q^2)$, allowing us to probe black holes of lower masses. Our calculations determine which interactions contribute to the two effects and how. This reveals what  ringdown signals can actually probe, and defines the simplest modelling framework for searching for massless scalars with black hole ringdowns. Our results are in agreement with the theory-agnostic statement of \cite{vdk4-tg61}\footnote{In \cite{vdk4-tg61, Li:2022pcy}, non-dynamical perturbations are parametrized by $\zeta \propto q^2$.}, namely that contamination generically appears at the same order in $q$ as the frequency shift, provided that the amplitude of the scalar perturbation is suppressed by $q$ (as assumed in \cite{vdk4-tg61}). When this assumption is dropped,  contamination can appear at order $q$ and hence dominate over frequency shifts. Our findings provide further motivation for going beyond the current black hole spectroscopy program by including the effects of frequency contamination and provide the framework for interpreting bounds in a theoretical context.

The rest of the paper is organized as follows. In Section \ref{sec:Theoretical_setup}, we give the action of shift-symmetric Horndeski theory and describe the perturbative formalism that we will use. This leads to the  calculation of the perturbed field equations in Section \ref{sec:Perturbation Equations}. We then use these to identify the individual terms contributing to the frequency shift and contamination in order to construct the ringdown ansatz in Section \ref{sec:Building_the_ringdown_ansatz}. Finally, we discuss the implications of these results on phenomenological modelling of bGR ringdowns and state our conclusions in Section \ref{sec:Discussion}. We use geometrized units where $\mathrm{G} = \mathrm{c} = 1$, unless otherwise specified. 

\section{\label{sec:Theoretical_setup}Theoretical setup}
\subsection{Action and field equations}\label{sec:action}
The action for shift-symmetric Horndeski gravity with a scalar field $\phi$ is 
\begin{equation}\label{eq:Horndeski Action}
S_{\text{H}}=\frac{1}{16 \pi} \sum_{i=2}^5 \int d^4 x \sqrt{-g} \ \mathcal{L}_i+S_{\text{M}},
\end{equation}
where the Horndeski sub-Lagrangians $\mathcal{L}_i$ are written as
\begin{equation}\label{eq:Horndeski Lagrangians}
\begin{aligned}
\mathcal{L}_2= & K(X) \\
\mathcal{L}_3= & -G_3(X) \Box \phi \\
\mathcal{L}_4= & G_4(X) R+G_{4 X}(X)\left[(\Box \phi)^2-\left(\nabla_\mu \nabla_\nu \phi\right)^2\right] \\
\mathcal{L}_5= & G_5(X) G_{\mu \nu} \nabla^\mu \nabla^\nu \phi-\frac{G_{5 X}}{6}\left[(\Box \phi)^3\right. \\
& \left.-3 \Box \phi\left(\nabla_\mu \nabla_\nu \phi\right)^2+2\left(\nabla_\mu \nabla_\nu \phi\right)^3\right],
\end{aligned}
\end{equation}
and $S_{\text{M}}$ is the matter action.
Here, $G_{\mu \nu}$ is the Einstein tensor, and $K$ and $G_i$ are some functions of the kinetic term $X \equiv -\frac{1}{2} \nabla_\mu \phi \nabla^\mu \phi$ \cite{PhysRevD.90.124063, 10.1143/PTP.126.511, Tanahashi_2017}, with derivatives denoted as $f_X\equiv df/dX$.Although $\mathcal{G}$ does not appear explicitly in \cref{eq:Horndeski Lagrangians}, it has been shown that a linear coupling $\phi \mathcal{G}$ can be generated by choosing $G_5(X) = - 4 \alpha \ln |X|$ \cite{10.1143/PTP.126.511}.

We will employ the approach of \cite{PhysRevD.109.084046} and push the calculation to second order in the charge per unit mass $q$. Thus, we wish to consider solutions of the field equations of the action \cref{eq:Horndeski Action} that are continuously connected to solutions of GR, while restricting ourselves to theories that respect local Lorentz symmetry (i.e. they admit constant $\phi$  for flat spacetime). The linear coupling to $\mathcal{G}$ is compatible with both conditions but it would seemingly contradict the latter when written as a contribution to  $G_5(X)$. To avoid this ambiguity, we  follow Ref.~ \cite{PhysRevD.99.124004} and first rewrite the Lagrangian as
\begin{equation}
\label{eq:ShiftSymmetricHorndeskiClassification}
    \mathcal{L} = \widetilde{ \mathcal{L}} + \alpha \phi \mathcal{G},
\end{equation}
and then impose that the (redefined) $G_i$ functions of $\widetilde{\mathcal{L}}$ can  be expanded as
\begin{equation}\label{eq:functions}
\begin{aligned}
\widetilde{K}(X) & =X+\mathcal{O}\left(X^2\right), \\
\widetilde{G}_3(X) & =\tau_3 X+\mathcal{O}\left(X^2\right), \\
\widetilde{G}_4(X) & =1+\tau_4 X+\mathcal{O}\left(X^2\right), \\
\widetilde{G}_5(X) & =\tau_5 X+\mathcal{O}\left(X^2\right).
\end{aligned}
\end{equation}
As we will confirm later, the $\order(X^2)$ terms do not contribute at the perturbative orders we will consider below.

The field equations for the metric derived from \cref{eq:ShiftSymmetricHorndeskiClassification,eq:functions} are
\begin{align}
\label{eq:FieldEqSum}
\sum_{i=2}^{5} F^i_{\mu\nu}
&+\frac{1}{2}\alpha
\Big[
\left(
g_{\rho\mu}g_{\delta\nu}
+g_{\rho\nu}g_{\delta\mu}
\right)
\nonumber\\
&\qquad\cdot
\nabla_\sigma\!\left(
\nabla_\gamma\phi\,
\varepsilon^{\lambda\eta\rho\sigma}
\varepsilon^{\alpha\beta\gamma\delta}
R_{\lambda\eta\alpha\beta}
\right)
\Big]
=0,
\end{align}
where $\varepsilon^{\lambda\eta\rho\sigma}$ is the Levi-Civita tensor.  The full expressions for the tensors $F^i_{\mu\nu}$ for general Horndeski theory are given in \cite{10.1143/PTP.126.511}, and the contribution from the $\alpha \phi \mathcal{G}$ coupling is found in \cite{PhysRevD.90.124063}.

 We assume that there is a single new scale in the theory and hence, in units where $G=c=1$, every coupling constant can be thought of as a dimensionless coefficient times a power of that scale. As discussed in more detail in Ref.~\cite{PhysRevD.109.084046}, using standard effective field theory arguments, this power can be determined by the dimensionality of each coupling after reinstating the Planck scale, with the expectation that the dimensionless coefficients are of order 1. Reexpressing the new scale in terms of $M$ then allows one to determine how the $\tau_i$ scale with $q$ \cite{PhysRevD.109.084046}: $\tau_3, \tau_4 \sim q$, and $\tau_5 \sim q^2$. As discussed recently in Ref.~\cite{thaalba2025screeningdipolaremissiontwoscale}, there can be theories with two disparate scales without compromising technical naturalness (in the quantum field theory sense). Such scenarios will not respect the aforementioned $\tau_i$ scaling with $q$ and we will not consider them here.

Turning now to the field equation for the scalar $\phi$, we note that the shift-symmetry of the scalar field brings with it a corresponding Noether current $\widetilde{J}^\mu$ as written in \cite{PhysRevD.90.124063}. Considering the linear coupling to the Gauss-Bonnet term, we then have the scalar equation
\begin{equation}\label{eq: EOM}
\nabla_\mu\left(\widetilde{J}^\mu-\alpha \mathcal{G}^\mu\right)=\nabla_\mu \widetilde{J}^\mu - \alpha \mathcal{G}=0 .
\end{equation}

\subsection{Perturbative formalism}\label{sec:perturbative formalism}
We will now perform a double expansion in the parameters $\epsilon$ and $q$. The first parameter, $\epsilon$, is associated with dynamical perturbations about a GR background, while $q$ is the scalar charge from \cref{eq:scalar charge} parameterizing the non-dynamical perturbations coming from the  bGR  terms.
We follow the notation from \cite{PhysRevD.109.084046} so that the order of a quantity with respect to $(\epsilon, q)$ is denoted by a superscript $(j,k)$. We will work at linear order in $\epsilon$ and quadratic order in $q$. The highest order terms we shall see are then of order $\epsilon q^2$, or $(j,k) = (1,2)$. Therefore, we are working at the same order as the theory-agnostic analysis in \cite{vdk4-tg61}  and \cite{Li:2022pcy}, and an order higher than the perturbative expansion in \cite{PhysRevD.109.084046}, which only went up to order $\epsilon q$. For simplicity, we denote background quantities where $(j,k)=(0,0)$ by an overbar. 

The perturbed metric $g_{\mu \nu}=\overline{g}_{\mu \nu}+h_{\mu \nu}$
is then 
\begin{equation}\label{eq:metric expansion}
\begin{aligned}
  g_{\mu \nu} &= \overline{g}_{\mu \nu}+\epsilon h_{\mu \nu}^{(1,0)} +q^2 h_{\mu \nu}^{(0,2)}+\epsilon q h_{\mu \nu}^{(1,1)}   + \epsilon q^2 h_{\mu \nu}^{(1,2)}  \\
  &= \overline{g}_{\mu \nu}+q^2 h_{\mu \nu}^{(0,2)} +\epsilon h^\mathrm{D}_{\mu \nu},
\end{aligned}    
\end{equation}
where $\overline{g}_{\mu \nu}$ corresponds to the background Kerr metric. In the second line, we have explicitly separated the dynamical part of the metric perturbation, $h^\mathrm{D}_{\mu \nu}$. 

Meanwhile, the scalar field is 
\begin{equation}\label{eq:scalar expansion}
    \phi=\epsilon \phi^{(1,0)}+ q \phi^{(0,1)}+q^2 \phi^{(0,2)}+\epsilon q \phi^{(1,1)} + \epsilon q^2\phi^{(1,2)}.
\end{equation}
As argued in \cite{PhysRevD.90.124063}, the stationary correction to the metric $qg_{\mu \nu}^{(0,1)}$ does not manifest physically, so we take $g_{\mu \nu}^{(0,1)} = 0$. This comes from the fact that in the stationary limit $\epsilon \rightarrow 0$, scalar perturbations first enter the metric as a backreaction, so they appear at order $q^2$. 

Meanwhile, assuming shift-symmetry means that we may unambiguously set the constant background scalar field to zero as $\overline{\phi} = 0$. 

It should be noted that the theory-agnostic analysis of \cite{vdk4-tg61} further assumes that $\phi^{(1,0)} = 0$, in order that $\phi \rightarrow 0$ as $q \rightarrow 0$. We will not make this  assumption for now and discuss it after we derive the perturbation equations and determine the most general ansatz for their solution.

\section{\label{sec:Perturbation Equations}Perturbation equations}

We now perturb the metric equation in \cref{eq:FieldEqSum}  and the scalar equation in \cref{eq: EOM} using our metric $g_{\mu \nu}$ from \cref{eq:metric expansion} and scalar $\phi$  from \cref{eq:scalar expansion}.

Solving \cref{eq: EOM} order by order, we find
\begin{align}
  \overline{\Box} \phi^{(1,0)}=& \ 0\; , \label{eq:BoxPhi10}\\
  \overline{\Box} \phi^{(0,1)}=&-\alpha^{(0,1)} \overline{R}^{ \mu \nu \alpha \beta} \overline{R}_{\mu \nu \alpha \beta} = -\alpha^{(0,1)}\overline{\mathcal{G}} \; ,
\label{eq:BoxPhi01}\\
\overline{\Box} \phi^{(0,2)}=& \ 0\; , \label{eq:BoxPhi02}\\ 
\overline{\Box} \phi^{(1,1)} =&-\alpha^{(0,1)}\left(2 \overline{R}^{\mu \nu \alpha \beta} R_{\mu \nu \alpha \beta}^{(1,0)} -h^{(1,0)}\overline{\mathcal{G}}\right) \nonumber \\
 &-\Box^{(1,0)}\phi^{(0,1)}
\label{eq:BoxPhi11}\; ,
\end{align}
where $h^{(1,0)} = \overline{g}^{\mu\nu}h^{(1,0)}_{\mu\nu}$.
To better keep track of perturbative  orders $(j,k)$, we have reinstated the relevant coupling constants as $\alpha \equiv q \alpha^{(0,1)}$ and $\tau_4 \equiv q \tau_4^{(0,1)}$. The perturbative approach leads to a key simplification; the operator acting on each perturbation is always the d'Alembertian defined on the background spacetime, $\overline \Box$.

Immediately, we can see from \cref{eq:BoxPhi10} that $\phi^{(1,0)}$ is a homogeneous solution. Meanwhile, the stationary equation for $\phi^{(0,1)}$ in \cref{eq:BoxPhi01} contains both the coupling $\alpha^{(0,1)}$ and the background Kretschmann scalar.

Next, considering that $\phi^{(0,2)}$ is another stationary correction to $\phi$, it is automatically time-independent. Combining this fact with the vanishing of \cref{eq:BoxPhi02}, it is clear from the no-hair theorem \cite{Hui:2012qt} that $\phi^{(0,2)}$ is a constant. Invoking shift-symmetry then, we may unambiguously set this to vanish as $\phi^{(0,2)}=0$. 

Now, the $\order(\epsilon q)$  \cref{eq:BoxPhi11} is not homogeneous. We see the dynamics of $\phi^{(1,1)}$ are determined by terms generated from the dynamical metric perturbation $h^{(1,0)}_{\mu \nu}$ along with non-dynamical scalar corrections. We omit the equation for $\phi^{(1,2)}$, as we shall see that it does not source anything in the field equations in \crefrange{eq:G10}{eq:G12} for the metric perturbations, at the orders we are considering.

We now turn to the perturbed field equations for the metric. We isolate the metric perturbations $h^{(j,k)}_{\mu \nu}$ from \cref{eq:FieldEqSum} by writing the linearized Einstein operator $\delta G_{\mu \nu}$ at each order. We then obtain a hierarchical set of field equations given by
\begin{widetext}
\begin{align}
  \delta G_{\mu\nu}[h^{(1, 0)}_{\mu\nu}]     =& \ 0\; , \label{eq:G10}\\
  \delta G_{\mu\nu}[h^{(0, 2)}_{\mu\nu}]   =&
 -\frac{1}{2}\alpha^{(0,1)}\left(\overline{g}_{\rho \mu} \overline{g}_{\delta \nu}+\overline{g}_{\rho \nu} \overline{g}_{\delta \mu}\right)
  \cdot \left[\overline{\nabla}_\sigma\left(\overline{\nabla}_\gamma \phi^{(0,1)} \overline{\varepsilon}^{\lambda \eta \rho \sigma } \overline{\varepsilon}^{ \alpha \beta \gamma \delta} \overline{R}_{\lambda \eta \alpha \beta}\right)\right] \notag\\
 & +\frac{1}{2} \overline{\nabla}_\mu \phi^{(0,1)} \overline{\nabla}_\nu \phi^{(0,1)}-\frac{1}{4} (\overline{\nabla}_\alpha \phi^{(0,1)})^2  \overline{g}_{\mu \nu} \; ,
\label{eq:G02}\\
\delta G_{\mu\nu}[h^{(1, 1)}_{\mu\nu}]  =& 
-\frac{1}{2}\alpha^{(0,1)}\left(\overline{g}_{\rho \mu}\overline{g}_{\delta \nu}+\overline{g}_{\rho \nu} \overline{g}_{\delta \mu}\right) \cdot \left[\overline{\nabla}_\sigma\left(\overline{\nabla}_\gamma \phi^{(1,0)} \overline{\varepsilon}^{ \lambda \eta \rho \sigma } \overline{\varepsilon}^{ \alpha \beta \gamma \delta} \overline{R}_{\lambda \eta \alpha \beta}\right)\right] \notag\\
& 
+\overline{\nabla}_{(\mu} \phi^{(1,0)} \overline{\nabla}_{\nu)} \phi^{(0,1)}-\frac{1}{2} \overline{\nabla}_\alpha \phi^{(1,0)} \overline{\nabla}^\alpha \phi^{(0,1)} \overline{g}_{\mu \nu}, \; \label{eq:G11}\\
\delta G_{\mu\nu}[h_{\mu\nu}^{(1,2)}] =& -\frac{1}{2}\alpha^{(0,1)} \Bigg\{
\Big(\overline{g}_{\rho\mu}h_{\delta\nu}^{(1,0)} + \overline{g}_{\rho\nu}h_{\delta\mu}^{(1,0)} + h_{\rho\mu}^{(1,0)}\overline{g}_{\delta\nu}  + h_{\rho\nu}^{(1,0)} \overline{g}_{\delta\mu}\Big)
\cdot\overline{\nabla}_\sigma \!\left(
\overline{\nabla}_\gamma \phi^{(0,1)}
\,\overline{\varepsilon}^{\lambda\eta\rho\sigma}
\,\overline{\varepsilon}^{\alpha\beta\gamma\delta}
\overline{R}_{\lambda\eta\alpha\beta}
\right) \notag
\\[2pt]
&+ \Big(\overline{g}_{\rho\mu} \overline{g}_{\delta\nu}
+ \overline{g}_{\rho\nu} \overline{g}_{\delta\mu}\Big)
\cdot \Bigg( \overline{\nabla}_\sigma \Big[
\left(\overline{\nabla}_\gamma \phi^{(1,1)}
\,\overline{\varepsilon}^{\lambda\eta\rho\sigma}
\,\overline{\varepsilon}^{\alpha\beta\gamma\delta}
\overline{R}_{\lambda\eta\alpha\beta}\right) \notag \\
&+ \left( \overline{\nabla}_\gamma \phi^{(0,1)}
\,\varepsilon^{(1,0)\lambda\eta\rho\sigma}
\,\overline{\varepsilon}^{\alpha\beta\gamma\delta}
\overline{R}_{\lambda\eta\alpha\beta}\right)
+ \left(\overline{\nabla}_\gamma \phi^{(0,1)}
\,\overline{\varepsilon}^{\lambda\eta\rho\sigma}
\,\varepsilon^{(1,0)\alpha\beta\gamma\delta}
\overline{R}_{\lambda\eta\alpha\beta}\right) \notag\\
&+ \left(\overline{\nabla}_\gamma \phi^{(0,1)}
\,\overline{\varepsilon}^{\lambda\eta\rho\sigma}
\,\overline{\varepsilon}^{\alpha\beta\gamma\delta}
R^{(1,0)}_{\lambda\eta\alpha\beta}\right)
\Big] + \nabla^{(1,0)}_\sigma\Big[\overline{\nabla}_\gamma \phi^{(0,1)}
\,\overline{\varepsilon}^{\lambda\eta\rho\sigma}
\,\overline{\varepsilon}^{\alpha\beta\gamma\delta}
\overline{R}_{\lambda\eta\alpha\beta}\Big] \Bigg)
\Bigg\}+ \overline{\nabla}_{(\mu} \phi^{(1,1)} \overline{\nabla}_{\nu)} \phi^{(0,1)} \notag\\[2pt]
&
- \frac{1}{2}\left(
\overline{\nabla}_\alpha \phi^{(1,1)} \overline{\nabla}^\alpha \phi^{(0,1)}\right) \overline{g}_{\mu\nu} - \frac{1}{4}
\Big(
\overline{\nabla}_\alpha \phi^{(0,1)}\overline{\nabla}^\alpha \phi^{(0,1)} 
\Big) h^{(1,0)}_{\mu\nu} +\frac{1}{4}\overline{g}_{\mu\nu}h^{(1,0)\alpha \beta}\overline{\nabla}_\alpha \phi^{(0,1)} \overline{\nabla}_\beta \phi^{(0,1)} \notag\\[2pt]
&+ \tau_4^{(0,1)} \Big\{
\overline{\Box} \phi^{(0,1)} \overline{\nabla}_\mu \overline{\nabla}_\nu \phi^{(1,0)}- 2 \,\overline{\nabla}_\lambda \overline{\nabla}_{(\mu} \phi^{(1,0)}
\,\overline{\nabla}^\lambda \overline{\nabla}_{\nu)} \phi^{(0,1)}+  \Big(\overline{\nabla}_\alpha \overline{\nabla}_\beta \phi^{(1,0)}
  \overline{\nabla}^\alpha \overline{\nabla}^\beta \phi^{(0,1)}
\Big)\overline{g}_{\mu\nu} \notag\\[2pt]
& \label{eq:G12}- 2 \overline{R}_{\mu\alpha\nu\beta}\overline{\nabla}^{(\alpha}\phi^{(1,0)}\overline{\nabla}^{\beta)}\phi^{(0,1)}  \Big\}-2\delta^2 G_{\mu 
\nu}[h^{(1,0)}_{\mu \nu},h^{(0,2)}_{\mu \nu}].  
\end{align}
\end{widetext}

We recover the results of \cite{PhysRevD.109.084046} up to $\order
(\epsilon q)$ in \crefrange{eq:G10}{eq:G11}. Although we have suppressed $\order(X^2)$ terms in \cref{eq:functions}, we have included them in the derivation of these equations and confirmed that they do not contribute at $\order(\epsilon q^2)$.
As was the case for $\overline{\Box}$ in the scalar sector, the operator acting on each metric perturbation $h_{\mu \nu}^{(j,k)}$ is a background quantity; here it is the linearized Einstein tensor, $\delta G_{\mu\nu}$. $\delta^2 G_{\mu\nu}$ is defined as the second order perturbation in $h_{\mu\nu}$ of the Einstein tensor \cite{PhysRevD.108.064002, PhysRevD.109.064022}. Given our expansion \cref{eq:metric expansion}, the only contribution of this operator to the equations up to $\order(\epsilon q^2)$ is$-2\delta^2 G_{\mu 
\nu}[h^{(1,0)}_{\mu \nu},h^{(0,2)}_{\mu \nu}]$.

We can see that \cref{eq:G10} is a homogeneous equation, and thus $h_{\mu \nu}^{(1,0)}$ only carries the purely GR Kerr QNMs \cite{PhysRevD.109.084046}. On the other hand,  it is evident from \cref{eq:G11} and \cref{eq:G12} that $h_{\mu \nu}^{(1,1)}$ and $h_{\mu \nu}^{(1,2)}$ carry bGR corrections to the QNMs.
Given that $h_{\mu \nu}^{(0,2)}$ is stationary by construction, we can see that \cref{eq:G02} is also purely stationary. However, we observe that $h_{\mu \nu}^{(0,2)}$ still contributes to the dynamical gravitational wave signal via the presence of the term $-2\delta^2 G_{\mu 
\nu}[h^{(1,0)}_{\mu \nu},h^{(0,2)}_{\mu \nu}]$ in \cref{eq:G12}.

\section{\label{sec:Building_the_ringdown_ansatz}Building the ringdown ansatz} 

In this section, we consider the hierarchical system of perturbed equations consisting of the scalar equations in \crefrange{eq:BoxPhi10}{eq:BoxPhi11} and the metric equations of \crefrange{eq:G10}{eq:G12}. This allows us to build an ansatz for the dynamical metric perturbation, focusing on the component of the ringdown described by QNMs as
\begin{equation}\label{eq:htilde simp}
 h^\mathrm{D}_{\mu \nu} =  h_{\mu \nu}^{(1,0)} + q h_{\mu \nu}^{(1,1)}   + q^2 h_{\mu \nu}^{(1,2)}
\end{equation}
step-by-step.

Beginning at $\order(\epsilon)$, the   QNM component of the  solution to the homogeneous differential equation in \cref{eq:G10} takes the form 
\begin{equation}\label{eq:h10}
    h^{(1,0)}_{\munu} \sim  \sum_\mathbf{n} A_\mathbf{n} \mathrm{e}^{-\mathrm{i} \omega^{(1,0)}_\mathbf{n}t},
\end{equation}
where $A_\mathbf{n}$ is an initial data dependent amplitude parameter given at a fixed spatial coordinate. For the remainder of the paper, we will use capital Latin letters with subscript $\mathbf{n}$ to denote similarly defined amplitude parameters for other perturbations. Furthermore, $\omega^{(1,0)}_\mathbf{n}$ are purely GR QNM frequencies, as in \cref{eq:GR QNM}.

Turning to the scalar equation at this order in \cref{eq:BoxPhi10}, we have a homogeneous equation for the scalar yielding a solution of the form 
\begin{equation}\label{eq:phi10}
    \phi^{(1,0)} \sim \sum_\mathbf{n} B_\mathbf{n} \mathrm{e}^{-\mathrm{i} \widehat{\omega}^{(1,0)}_\mathbf{n}t}.
\end{equation}
 Here, we use the $\widehat{\omega}^{(j,k)}_\mathbf{n}$ to denote scalar frequencies, leaving $\omega^{(j,k)}_\mathbf{n}$ for frequencies associated with the metric. Importantly, these frequencies are generically going to be distinct for two separate fields , such as metric and scalar, because the operators acting on them are different. 

Moving up now to $\order(\epsilon q)$, we see that the dynamical part of \cref{eq:G11} comes entirely from $\phi^{(1,0)}$. Therefore, its general solution $h_{\mu \nu}^{(1,1)}$ is formed of both the homogeneous solution in \cref{eq:h10} and the particular solution coming entirely from QNMs of the scalar field $\phi^{(1,0)}$ in \cref{eq:phi10}. This perturbation then reads as
\begin{equation}\label{eq:h11}
    h^{(1,1)}_{\munu} \sim \sum_\mathbf{n} \Big[ C_\mathbf{n} \mathrm{e}^{-\mathrm{i} \omega^{(1,0)}_\mathbf{n}t} + D_\mathbf{n} \mathrm{e}^{-\mathrm{i} \widehat{\omega}^{(1,0)}_\mathbf{n}t}\Big].
\end{equation}
Thus, we can see that the bGR correction to $h^{(1,1)}_{\munu}$ comes entirely from scalar contamination, and has no contribution from a frequency shift of the form of \cref{eq:Shift QNM}.

Looking now to the $\phi^{(1,1)}$ equation in \cref{eq:BoxPhi11}, we see that it contains terms dependent upon the dynamical metric term $h_{\mu \nu}^{(1,0)}$. Therefore, its solution $\phi^{(1,1)}$ carries both GR and bGR scalar QNMs as 
\begin{equation}\label{eq:phi11}
    \phi^{(1,1)} \sim \sum_\mathbf{n} \Big[ E_\mathbf{n}\mathrm{e}^{-\mathrm{i} \widehat{\omega}^{(1,0)}_\mathbf{n}t} + F_\mathbf{n}\mathrm{e}^{-\mathrm{i} \omega^{(1,0)}_\mathbf{n}t}\Big].
\end{equation}

Finally, at $\order(\epsilon q^2)$, the highest order we consider, we find that the dynamics of \cref{eq:G11} are governed by terms containing $h_{\mu \nu}^{(1,0)}$, $\phi^{(1,0)}$, and $\phi^{(1,1)}$. We can then rewrite this equation as
\begin{equation}\label{eq:G12 Split}
    \delta G_{\mu\nu}[h_{\mu\nu}^{(1,2)}] = S_{\mu\nu}^{(1,2)}[h_{\mu\nu}^{(1,0)}] + S_{\mu\nu}^{(1,2)}[\phi^{(1,0)}] + S_{\mu\nu}^{(1,2)}[\phi^{(1,1)}]. 
\end{equation}
We must then determine how all these dynamical terms contribute to the solution $h_{\mu \nu}^{(1,2)}$ of \cref{eq:G11}.

We first look at $S_{\mu\nu}^{(1,2)}[h_{\mu\nu}^{(1,0)}]$. As the dynamics are sourced by $h_{\mu \nu}^{(1,0)}$, we have a shift from the GR QNM frequencies  $\omega^{(1,0)}_\mathbf{n}$  as 
\begin{equation}
    \omega^{(1,0)}_\mathbf{n} \rightarrow \omega^{(1,0)}_\mathbf{n} + q^2 \omega^{(1,2)}_\mathbf{n},
    \label{eq:FrequencyShift}
\end{equation}
where $\delta \omega_\mathbf{n} = q^2 \omega^{(1,2)}_\mathbf{n}$ as modelled in the notation of \cref{eq:Shift QNM} \cite{Li:2023ulk}.
Apart from the frequency shift in  \cref{eq:FrequencyShift}, 
we should expect an additional shift of the spatial mode function 
$u_\mathbf{n}(x^i)$ ,
\begin{equation}
   u^{(0)}_\mathbf{n}(x^i) \mathrm{e}^{-\mathrm{i} \omega^{(1,0)}_\mathbf{n}t} \rightarrow \left(u^{(0)}_\mathbf{n}(x^i)+q^2u^{(2)}_\mathbf{n}(x^i)\right)\mathrm{e}^{-\mathrm{i} \omega^{(1,0)}_\mathbf{n} t},
\label{eq:RadialShift}
\end{equation}
where $u^{(k)}_\mathbf{n}$ denotes the order of the spatial mode function with respect to $q$.
By taking our ansatz to be evaluated at null infinity, however, we can absorb this spatial shift effect into the amplitudes  \cite{vdk4-tg61}.

Next, with $S_{\mu\nu}^{(1,2)}[\phi^{(1,0)}]$, the dynamical part carries the scalar QNM frequencies $\widehat{\omega}^{(1,0)}_\mathbf{n}$ as in \cref{eq:phi10}.
Lastly, dealing with $S_{\mu\nu}^{(1,2)}[\phi^{(1,1)}]$ is the most complicated step, given that $\phi^{(1,1)}$ in \cref{eq:phi11} contains both the GR QNM frequencies $\omega^{(1,0)}_\mathbf{n}$ and the scalar QNM frequencies $\widehat{\omega}^{(1,0)}_\mathbf{n}$. This means that the $\phi^{(1,1)}$  contributes to both the frequency shift and the scalar contamination of $h_{\mu \nu}^{(1,2)}$ \cite{vdk4-tg61}.

We then have

\begin{equation}\label{eq:h12}
    h^{(1,2)}_{\munu} \sim \sum_\mathbf{n} \Big[ G_\mathbf{n}\mathrm{e}^{-\mathrm{i} (\omega^{(1,0)}_\mathbf{n}+q^2 \omega^{(1,2)}_\mathbf{n})t} + H_\mathbf{n}\mathrm{e}^{-\mathrm{i} \widehat{\omega}^{(1,0)}_\mathbf{n}t}\Big],  
\end{equation}
which makes clear that at $\order(\epsilon q^2)$, we have bGR corrections from both the shift and contamination.

Putting \cref{eq:h10}, \cref{eq:h11}, and \cref{eq:h12} together into \cref{eq:htilde simp}, we get, in summary
\begin{align}\label{eq:htilde full simp}
    h^\mathrm{D}_{\munu} =  \sum_\mathbf{n} \Big[\widetilde{A}_\mathbf{n} & \mathrm{e}^{-\mathrm{i}( \omega^{(1,0)}_\mathbf{n}+q^2\omega^{(1,2)}_\mathbf{n} )t} \notag \\
    & + \left(q D_\mathbf{n}+ q^2 H_\mathbf{n}\right)\mathrm{e}^{-\mathrm{i} \widehat{\omega}^{(1,0)}_\mathbf{n}t} \Big].
\end{align}
Here, we have relabelled the coefficient of the shift term to $\widetilde{A}_\mathbf{n}$, since we can group together terms with the same time dependence. We have also absorbed the spatial shift effect from \cref{eq:RadialShift} into $\widetilde{A}_\mathbf{n}$. 

This clearly shows which terms contribute to the dynamical metric perturbation at each order. Starting at $\order(\epsilon)$, we simply have the GR QNM frequencies $\omega^{(1,0)}_\mathbf{n}$. At $\order(\epsilon q)$, the only bGR correction is from the contamination term $qD_\mathbf{n}\mathrm{e}^{-\mathrm{i} \widehat{\omega}^{(1,0)}_\mathbf{n}t}$. Lastly, at $\order(\epsilon q^2)$, we have both the GR frequencies $\omega^{(1,0)}_\mathbf{n}$ being modified by the shift $\delta \omega_\mathbf{n}=q^2\omega^{(1,2)}_\mathbf{n}$ in the exponent along with scalar contamination coming from the $q^2H_\mathbf{n}\mathrm{e}^{-\mathrm{i} \widehat{\omega}^{(1,0)}_\mathbf{n}t}$  correction. 
Note that as $q \rightarrow 0$ our ansatz reduces to the one for a GR ringdown.

It is worth stressing that  contamination by the scalar frequencies first arises at $\order(\epsilon q)$. Hence, there might be scenarios in which it dominates over the frequency shifts, which first appear at $\order(\epsilon q^2)$.

\subsection{Sources of the contamination and shift}
We have found that bGR corrections to the field equations come in at $\order(\epsilon q)$ and $\order(\epsilon q^2)$. Starting at $\order(\epsilon q)$, and considering $\delta G_{\mu\nu}[h^{(1, 1)}_{\mu\nu}]$ in \cref{eq:G11}, we see that we only have terms arising from the linear coupling $\alpha \phi\mathcal{ G}$, as found by \cite{PhysRevD.109.084046}. As we have shown in \cref{eq:h11}, the only bGR contribution to $h^{(1,1)}_{\mu \nu}$ is from scalar contamination. This then means that the $qD_\mathbf{n}\mathrm{e}^{-\mathrm{i} \widehat{\omega}^{(1,0)}_\mathbf{n}t}$ component of $h^{(1,2)}_{\mu \nu}$ in \cref{eq:htilde full simp} comes solely from this coupling. 

Now, at $\order(\epsilon q^2)$, we have both scalar contamination and frequency shift. Comparing \cref{eq:G12} and \cref{eq:G12 Split}, we see that the frequency shift contributions from $S_{\mu\nu}^{(1,2)}[h_{\mu\nu}^{(1,0)}]$ and $S_{\mu\nu}^{(1,2)}[\phi^{(1,1)}]$ come entirely from the $\alpha \phi \mathcal{G}$ coupling.

Likewise, the sourcing of scalar contamination from $S_{\mu\nu}^{(1,2)}[\phi^{(1,1)}]$ is exclusively from the same  $\alpha \phi \mathcal{G}$ coupling.

We now turn our attention to the term $S_{\mu\nu}^{(1,2)}[\phi^{(1,0)}]$, which  contributes purely to scalar contamination. We can see from \cref{eq:G12} that it contains terms coming from both the $\alpha \phi \mathcal{G}$ coupling and the quartic Horndeski $\mathcal{L}_4$ sector of \cref{eq:Horndeski Lagrangians} with its coupling constant $\tau_4$. Thus, when we look at the ansatz for $ h^{(1,2)}_{\munu}$ in \cref{eq:h12}, the shift term $G\mathrm{e}^{-\mathrm{i} (\omega^{(1,0)}_\mathbf{n}+q^2 \omega^{(1,2)}_\mathbf{n})t}$ only contains information about the $\alpha \phi \mathcal{G}$ coupling, while the contamination term $H_\mathbf{n}\mathrm{e}^{-\mathrm{i} \widehat{\omega}^{(1,0)}_\mathbf{n}t}$ is generated by both the $\alpha \phi \mathcal{G}$ coupling and the $\tau_4$ coupling. 

Taking it all together, the amplitudes $\widetilde{A}_\mathbf{n}$ and $qD_\mathbf{n}$ of $h^\mathrm{D}_{\munu}$ in \cref{eq:htilde full simp} contain information only about $\alpha$. Meanwhile, the coefficient $q^2H_\mathbf{n}$ carries knowledge of some combination of both $\alpha$ and $\tau_4$. While in practice, these amplitudes would also take input from initial data, here, we have limited the analysis to identifying the independent couplings appearing at each order. We note that $\tau_3$ and $\tau_5$ couplings from the $\mathcal{L}_3$ and $\mathcal{L}_5$ Horndeski sectors do not contribute to the field equations at these orders.

\section{\label{sec:Discussion}Discussion and Conclusions}

We have modelled the ringdown of a hairy black hole in shift-symmetric Horndeski gravity. We have exploited the fact that the scalar charge is not an independent parameter, but is instead determined by the black hole mass (and potentially spin) and the coupling constants of the theory, with the linear coupling between the scalar and the Gauss-Bonnet invariant being the dominant contribution to the charge. Assuming that there is a single new scale in the theory, we were then able to set up a perturbative scheme in the charge per unit mass $q$ and work at linear order in the perturbations of the fields, with bookkeeping parameter $\epsilon$. 

We have derived perturbation equations and the most general ansatz for their solution at $\order(\epsilon q^2)$. This extends the analysis of \cite{PhysRevD.109.084046} and reaches the same perturbative order as the theory-agnostic ansatz derived in \cite{vdk4-tg61}, thereby allowing us to determine which terms in the shift-symmetric Horndeski action contribute to deviations from GR. 

Our results reveal that only two coupling constants are relevant up to $\order(\epsilon q^2)$, $\alpha$ and $\tau_4$. These are associated with the  $\phi {\cal G}$ interaction that introduces hair and with the leading order contribution of the $\tilde{G}_4$ term respectively. This demonstrates that, in the small charge limit, the effects of a massless scalar on black hole ringdowns can be captured by a fairly simple effective field theory, without the need for  theory-agnostic or parametrized frameworks or approximations. Considering that $q$ is already constrained to be below 10\% \cite{Julie:2024fwy, PhysRevD.110.044022, r43k-51yq,Maselli:2020zgv,Maselli:2021men,Speri:2024qak} for solar mass black holes and scales inversely proportional to the  black hole mass squared, $q \propto M^{-2}$,
working at $\order(\epsilon q^2)$ can prove to be an adequate approximation in most cases, especially if future bounds from comparable mass or EMRI inspirals become even tighter.

Certain terms in the equations, and consequently in the ringdown modelling ansatz, depend on $\phi^{(1,0)}$ --- the contribution to the scalar amplitude as $q\to 0$ --- and vanish when $\phi^{(1,0)}=0$. In this case, both the frequency shift and contamination appear and depend only on $\alpha$. That is, including the $\phi {\cal G}$ interaction suffices to model ringdown fully at order $\order(\epsilon q^2)$ and  contamination cannot be neglected with respect to frequency shifts. 
 This is consistent with the theory-agnostic analysis of \cite{vdk4-tg61}, which indeed assumed that $\phi^{(1,0)}=0$.

The 
latter is a common assumption \cite{vdk4-tg61,Li:2022pcy} and it is the reasonable choice in scenarios where the scalar amplitude is expected to be naturally suppressed by $q$. If instead the scalar has a significantly larger amplitude than the metric perturbations, then the $\phi^{(1,0)}=0$ assumption might no longer be justified in a ringdown modelling scenario \cite{PhysRevD.109.084046,vdk4-tg61}. In that case, contamination first appears  at $\order(\epsilon q)$, while frequency shifts first appear at $\order(\epsilon q^2)$.  This suggests that in scenarios where, e.g.~nonlinearities during the merger could amplify the scalar field, contamination could become the dominant observable effect. It is not clear if massless (shift-symmetric) scalars can exhibit such behaviour. Such nonlinear effects are more likely  to appear in models that exhibit spontaneous scalarization \cite{Silva:2017uqg, Doneva:2017bvd, Ventagli:2020rnx,PhysRevLett.125.231101, PhysRevLett.126.011103,RevModPhys.96.015004}. It would be very interesting to extend our analysis to non-shift-symmetric theories. Future work using numerical relativity simulations of mergers could potentially clarify if taking $\phi^{(1,0)}\neq0$ is generally well motivated or if there are interesting exceptions.

Our results provide further support for the conclusion of Ref.~\cite{PhysRevD.109.084046} that LISA ringdowns from comparable mass binaries of supermassive black holes are not a good probe for massless scalar fields due to the scaling of the scalar with the black hole mass.  However, the Einstein Telescope \cite{Punturo_2010, Abac_2026} and Cosmic Explorer \cite{2019BAAS...51g..35R, Evans:2021gyd} will be able to provide more stringent tests of both frequency shifts and contamination for solar mass black holes. Our results provide an efficient way to model ringdowns, enable searches for both effects, and translate  theory-agnostic inference  \cite{SciPostPhys.20.1.025, ptg5-f769} into bounds on universal coupling constants that control the size of these effects.

\begin{acknowledgments}

We would like to thank Reinosuke Kusano, Jacopo Lestingi, and Lorenzo Pompili for helpful comments on the manuscript. TPS acknowledges partial support from the STFC Consolidated Grant nos. ST/V005596/1,  ST/X000672/1, and UKRI2492.

\end{acknowledgments}

\bibliography{references}

\begin{thebibliography}{80}%
\makeatletter
\providecommand \@ifxundefined [1]{%
 \@ifx{#1\undefined}
}%
\providecommand \@ifnum [1]{%
 \ifnum #1\expandafter \@firstoftwo
 \else \expandafter \@secondoftwo
 \fi
}%
\providecommand \@ifx [1]{%
 \ifx #1\expandafter \@firstoftwo
 \else \expandafter \@secondoftwo
 \fi
}%
\providecommand \natexlab [1]{#1}%
\providecommand \enquote  [1]{``#1''}%
\providecommand \bibnamefont  [1]{#1}%
\providecommand \bibfnamefont [1]{#1}%
\providecommand \citenamefont [1]{#1}%
\providecommand \href@noop [0]{\@secondoftwo}%
\providecommand \href [0]{\begingroup \@sanitize@url \@href}%
\providecommand \@href[1]{\@@startlink{#1}\@@href}%
\providecommand \@@href[1]{\endgroup#1\@@endlink}%
\providecommand \@sanitize@url [0]{\catcode `\\12\catcode `\$12\catcode `\&12\catcode `\#12\catcode `\^12\catcode `\_12\catcode `\%12\relax}%
\providecommand \@@startlink[1]{}%
\providecommand \@@endlink[0]{}%
\providecommand \url  [0]{\begingroup\@sanitize@url \@url }%
\providecommand \@url [1]{\endgroup\@href {#1}{\urlprefix }}%
\providecommand \urlprefix  [0]{URL }%
\providecommand \Eprint [0]{\href }%
\providecommand \doibase [0]{http://dx.doi.org/}%
\providecommand \selectlanguage [0]{\@gobble}%
\providecommand \bibinfo  [0]{\@secondoftwo}%
\providecommand \bibfield  [0]{\@secondoftwo}%
\providecommand \translation [1]{[#1]}%
\providecommand \BibitemOpen [0]{}%
\providecommand \bibitemStop [0]{}%
\providecommand \bibitemNoStop [0]{.\EOS\space}%
\providecommand \EOS [0]{\spacefactor3000\relax}%
\providecommand \BibitemShut  [1]{\csname bibitem#1\endcsname}%
\let\auto@bib@innerbib\@empty
\bibitem [{\citenamefont {Clemence}(1947)}]{RevModPhys.19.361}%
  \BibitemOpen
  \bibfield  {author} {\bibinfo {author} {\bibfnamefont {G.~M.}\ \bibnamefont {Clemence}},\ }\href {\doibase 10.1103/RevModPhys.19.361} {\bibfield  {journal} {\bibinfo  {journal} {Rev. Mod. Phys.}\ }\textbf {\bibinfo {volume} {19}},\ \bibinfo {pages} {361} (\bibinfo {year} {1947})}\BibitemShut {NoStop}%
\bibitem [{\citenamefont {Gilmore}\ and\ \citenamefont {Tausch-Pebody}(2021)}]{Gilmore_2021}%
  \BibitemOpen
  \bibfield  {author} {\bibinfo {author} {\bibfnamefont {G.}~\bibnamefont {Gilmore}}\ and\ \bibinfo {author} {\bibfnamefont {G.}~\bibnamefont {Tausch-Pebody}},\ }\href {\doibase 10.1098/rsnr.2020.0040} {\bibfield  {journal} {\bibinfo  {journal} {Notes Rec.}\ }\textbf {\bibinfo {volume} {76}},\ \bibinfo {pages} {155–180} (\bibinfo {year} {2021})}\BibitemShut {NoStop}%
\bibitem [{\citenamefont {Pound}\ and\ \citenamefont {Rebka}(1960)}]{PhysRevLett.4.337}%
  \BibitemOpen
  \bibfield  {author} {\bibinfo {author} {\bibfnamefont {R.~V.}\ \bibnamefont {Pound}}\ and\ \bibinfo {author} {\bibfnamefont {G.~A.}\ \bibnamefont {Rebka}},\ }\href {\doibase 10.1103/PhysRevLett.4.337} {\bibfield  {journal} {\bibinfo  {journal} {Phys. Rev. Lett.}\ }\textbf {\bibinfo {volume} {4}},\ \bibinfo {pages} {337} (\bibinfo {year} {1960})}\BibitemShut {NoStop}%
\bibitem [{\citenamefont {Hafele}\ and\ \citenamefont {Keating}(1972)}]{1972Sci...177..166H}%
  \BibitemOpen
  \bibfield  {author} {\bibinfo {author} {\bibfnamefont {J.~C.}\ \bibnamefont {Hafele}}\ and\ \bibinfo {author} {\bibfnamefont {R.~E.}\ \bibnamefont {Keating}},\ }\href {\doibase 10.1126/science.177.4044.166} {\bibfield  {journal} {\bibinfo  {journal} {Science}\ }\textbf {\bibinfo {volume} {177}},\ \bibinfo {pages} {166} (\bibinfo {year} {1972})}\BibitemShut {NoStop}%
\bibitem [{\citenamefont {{Hulse}}\ and\ \citenamefont {{Taylor}}(1975)}]{1975ApJ...195L..51H}%
  \BibitemOpen
  \bibfield  {author} {\bibinfo {author} {\bibfnamefont {R.~A.}\ \bibnamefont {{Hulse}}}\ and\ \bibinfo {author} {\bibfnamefont {J.~H.}\ \bibnamefont {{Taylor}}},\ }\href {\doibase 10.1086/181708} {\bibfield  {journal} {\bibinfo  {journal} {Astrophys. J. Lett.}\ }\textbf {\bibinfo {volume} {195}},\ \bibinfo {pages} {L51} (\bibinfo {year} {1975})}\BibitemShut {NoStop}%
\bibitem [{\citenamefont {Everitt}\ \emph {et~al.}(2011)\citenamefont {Everitt} \emph {et~al.}}]{PhysRevLett.106.221101}%
  \BibitemOpen
  \bibfield  {author} {\bibinfo {author} {\bibfnamefont {C.~W.~F.}\ \bibnamefont {Everitt}} \emph {et~al.},\ }\href {\doibase 10.1103/PhysRevLett.106.221101} {\bibfield  {journal} {\bibinfo  {journal} {Phys. Rev. Lett.}\ }\textbf {\bibinfo {volume} {106}},\ \bibinfo {pages} {221101} (\bibinfo {year} {2011})}\BibitemShut {NoStop}%
\bibitem [{\citenamefont {Abbot}\ \emph {et~al.}(2016)\citenamefont {Abbot} \emph {et~al.}}]{Abbott_2016}%
  \BibitemOpen
  \bibfield  {author} {\bibinfo {author} {\bibfnamefont {B.}~\bibnamefont {Abbot}} \emph {et~al.} (\bibinfo {collaboration} {LIGO Scientific Collaboration and the Virgo Collaboration}),\ }\href {\doibase 10.1103/PhysRevLett.116.061102} {\bibfield  {journal} {\bibinfo  {journal} {Phys. Rev. Lett.}\ }\textbf {\bibinfo {volume} {116}},\ \bibinfo {eid} {061102} (\bibinfo {year} {2016})}\BibitemShut {NoStop}%
\bibitem [{\citenamefont {Abbott}\ \emph {et~al.}(2021)\citenamefont {Abbott} \emph {et~al.}}]{Abbott_2021}%
  \BibitemOpen
  \bibfield  {author} {\bibinfo {author} {\bibfnamefont {R.}~\bibnamefont {Abbott}} \emph {et~al.} (\bibinfo {collaboration} {LIGO Scientific Collaboration and the Virgo Collaboration}),\ }\href {\doibase 10.1103/PhysRevX.11.021053} {\bibfield  {journal} {\bibinfo  {journal} {Phys. Rev. X}\ }\textbf {\bibinfo {volume} {11}},\ \bibinfo {pages} {021053} (\bibinfo {year} {2021})}\BibitemShut {NoStop}%
\bibitem [{\citenamefont {Abac}\ \emph {et~al.}(2025{\natexlab{a}})\citenamefont {Abac} \emph {et~al.}}]{Abac_2025}%
  \BibitemOpen
  \bibfield  {author} {\bibinfo {author} {\bibfnamefont {A.~G.}\ \bibnamefont {Abac}} \emph {et~al.} (\bibinfo {collaboration} {LIGO Scientific Collaboration, the Virgo Collaboration, and the KAGRA Collaboration}),\ }\href {\doibase 10.3847/2041-8213/ae0c06} {\bibfield  {journal} {\bibinfo  {journal} {The Astrophysical Journal Letters}\ }\textbf {\bibinfo {volume} {995}},\ \bibinfo {pages} {L18} (\bibinfo {year} {2025}{\natexlab{a}})}\BibitemShut {NoStop}%
\bibitem [{\citenamefont {Abbott}\ \emph {et~al.}(2016)\citenamefont {Abbott} \emph {et~al.}}]{PhysRevLett.116.221101}%
  \BibitemOpen
  \bibfield  {author} {\bibinfo {author} {\bibfnamefont {B.~P.}\ \bibnamefont {Abbott}} \emph {et~al.} (\bibinfo {collaboration} {LIGO Scientific Collaboration and the Virgo Collaboration}),\ }\href {\doibase 10.1103/PhysRevLett.116.221101} {\bibfield  {journal} {\bibinfo  {journal} {Phys. Rev. Lett.}\ }\textbf {\bibinfo {volume} {116}},\ \bibinfo {pages} {221101} (\bibinfo {year} {2016})}\BibitemShut {NoStop}%
\bibitem [{\citenamefont {Abbott}\ \emph {et~al.}(2019)\citenamefont {Abbott} \emph {et~al.}}]{Abbott_2019}%
  \BibitemOpen
  \bibfield  {author} {\bibinfo {author} {\bibfnamefont {B.~P.}\ \bibnamefont {Abbott}} \emph {et~al.} (\bibinfo {collaboration} {LIGO Scientific Collaboration and the Virgo Collaboration}),\ }\href {\doibase 10.1103/PhysRevLett.123.011102} {\bibfield  {journal} {\bibinfo  {journal} {Phys. Rev. Lett.}\ }\textbf {\bibinfo {volume} {123}},\ \bibinfo {pages} {011102} (\bibinfo {year} {2019})}\BibitemShut {NoStop}%
\bibitem [{\citenamefont {Abac}\ \emph {et~al.}(2026{\natexlab{a}})\citenamefont {Abac} \emph {et~al.}}]{6c61-fm1n}%
  \BibitemOpen
  \bibfield  {author} {\bibinfo {author} {\bibfnamefont {A.~G.}\ \bibnamefont {Abac}} \emph {et~al.} (\bibinfo {collaboration} {LIGO Scientific Collaboration, The Virgo Collaboration, and The KAGRA Collaboration}),\ }\href {\doibase 10.1103/6c61-fm1n} {\bibfield  {journal} {\bibinfo  {journal} {Phys. Rev. Lett.}\ }\textbf {\bibinfo {volume} {136}},\ \bibinfo {pages} {041403} (\bibinfo {year} {2026}{\natexlab{a}})}\BibitemShut {NoStop}%
\bibitem [{\citenamefont {{Punturo}}\ \emph {et~al.}(2010)\citenamefont {{Punturo}} \emph {et~al.}}]{Punturo_2010}%
  \BibitemOpen
  \bibfield  {author} {\bibinfo {author} {\bibfnamefont {M.}~\bibnamefont {{Punturo}}} \emph {et~al.},\ }\href {\doibase 10.1088/0264-9381/27/19/194002} {\bibfield  {journal} {\bibinfo  {journal} {Class. Quantum Grav.}\ }\textbf {\bibinfo {volume} {27}},\ \bibinfo {eid} {194002} (\bibinfo {year} {2010})}\BibitemShut {NoStop}%
\bibitem [{\citenamefont {Abac}\ \emph {et~al.}(2026{\natexlab{b}})\citenamefont {Abac} \emph {et~al.}}]{Abac_2026}%
  \BibitemOpen
  \bibfield  {author} {\bibinfo {author} {\bibfnamefont {A.}~\bibnamefont {Abac}} \emph {et~al.} (\bibinfo {collaboration} {Einstein Telescope Collaboration}),\ }\href {\doibase 10.1088/1475-7516/2026/03/081} {\bibfield  {journal} {\bibinfo  {journal} {J. Cosmol. Astropart. Phys.}\ }\textbf {\bibinfo {volume} {2026}},\ \bibinfo {pages} {081} (\bibinfo {year} {2026}{\natexlab{b}})}\BibitemShut {NoStop}%
\bibitem [{\citenamefont {{Reitze}}\ \emph {et~al.}(2019)\citenamefont {{Reitze}} \emph {et~al.}}]{2019BAAS...51g..35R}%
  \BibitemOpen
  \bibfield  {author} {\bibinfo {author} {\bibfnamefont {D.}~\bibnamefont {{Reitze}}} \emph {et~al.},\ }in\ \href {\doibase 10.48550/arXiv.1907.04833} {\emph {\bibinfo {booktitle} {Bulletin of the American Astronomical Society}}},\ Vol.~\bibinfo {volume} {51}\ (\bibinfo {year} {2019})\ p.~\bibinfo {pages} {35}\BibitemShut {NoStop}%
\bibitem [{\citenamefont {Evans}\ \emph {et~al.}(2021)\citenamefont {Evans} \emph {et~al.}}]{Evans:2021gyd}%
  \BibitemOpen
  \bibfield  {author} {\bibinfo {author} {\bibfnamefont {M.}~\bibnamefont {Evans}} \emph {et~al.},\ }\href@noop {} {\  (\bibinfo {year} {2021})},\ \Eprint {http://arxiv.org/abs/2109.09882} {arXiv:2109.09882 [astro-ph.IM]} \BibitemShut {NoStop}%
\bibitem [{\citenamefont {Amaro-Seoane}\ \emph {et~al.}(2017)\citenamefont {Amaro-Seoane} \emph {et~al.}}]{amaroseoane2017laserinterferometerspaceantenna}%
  \BibitemOpen
  \bibfield  {author} {\bibinfo {author} {\bibfnamefont {P.}~\bibnamefont {Amaro-Seoane}} \emph {et~al.} (\bibinfo {collaboration} {LISA Collaboration}),\ }\href {https://arxiv.org/abs/1702.00786} {\  (\bibinfo {year} {2017})},\ \Eprint {http://arxiv.org/abs/1702.00786} {arXiv:1702.00786} \BibitemShut {NoStop}%
\bibitem [{\citenamefont {Israel}(1967)}]{PhysRev.164.1776}%
  \BibitemOpen
  \bibfield  {author} {\bibinfo {author} {\bibfnamefont {W.}~\bibnamefont {Israel}},\ }\href {\doibase 10.1103/PhysRev.164.1776} {\bibfield  {journal} {\bibinfo  {journal} {Phys. Rev.}\ }\textbf {\bibinfo {volume} {164}},\ \bibinfo {pages} {1776} (\bibinfo {year} {1967})}\BibitemShut {NoStop}%
\bibitem [{\citenamefont {Carter}(1971)}]{PhysRevLett.26.331}%
  \BibitemOpen
  \bibfield  {author} {\bibinfo {author} {\bibfnamefont {B.}~\bibnamefont {Carter}},\ }\href {\doibase 10.1103/PhysRevLett.26.331} {\bibfield  {journal} {\bibinfo  {journal} {Phys. Rev. Lett.}\ }\textbf {\bibinfo {volume} {26}},\ \bibinfo {pages} {331} (\bibinfo {year} {1971})}\BibitemShut {NoStop}%
\bibitem [{\citenamefont {Robinson}(1975)}]{PhysRevLett.34.905}%
  \BibitemOpen
  \bibfield  {author} {\bibinfo {author} {\bibfnamefont {D.~C.}\ \bibnamefont {Robinson}},\ }\href {\doibase 10.1103/PhysRevLett.34.905} {\bibfield  {journal} {\bibinfo  {journal} {Phys. Rev. Lett.}\ }\textbf {\bibinfo {volume} {34}},\ \bibinfo {pages} {905} (\bibinfo {year} {1975})}\BibitemShut {NoStop}%
\bibitem [{\citenamefont {Kerr}(1963)}]{PhysRevLett.11.237}%
  \BibitemOpen
  \bibfield  {author} {\bibinfo {author} {\bibfnamefont {R.~P.}\ \bibnamefont {Kerr}},\ }\href {\doibase 10.1103/PhysRevLett.11.237} {\bibfield  {journal} {\bibinfo  {journal} {Phys. Rev. Lett.}\ }\textbf {\bibinfo {volume} {11}},\ \bibinfo {pages} {237} (\bibinfo {year} {1963})}\BibitemShut {NoStop}%
\bibitem [{\citenamefont {Leaver}(1985)}]{10.1098/rspa.1985.0119}%
  \BibitemOpen
  \bibfield  {author} {\bibinfo {author} {\bibfnamefont {E.~W.}\ \bibnamefont {Leaver}},\ }\href {\doibase 10.1098/rspa.1985.0119} {\bibfield  {journal} {\bibinfo  {journal} {Proc. Roy. Soc. Lond. A}\ }\textbf {\bibinfo {volume} {402}},\ \bibinfo {pages} {285} (\bibinfo {year} {1985})}\BibitemShut {NoStop}%
\bibitem [{\citenamefont {Nollert}(1999)}]{Hans-Peter_Nollert_1999}%
  \BibitemOpen
  \bibfield  {author} {\bibinfo {author} {\bibfnamefont {H.-P.}\ \bibnamefont {Nollert}},\ }\href {\doibase 10.1088/0264-9381/16/12/201} {\bibfield  {journal} {\bibinfo  {journal} {Class. Quantum Grav.}\ }\textbf {\bibinfo {volume} {16}},\ \bibinfo {pages} {R159} (\bibinfo {year} {1999})}\BibitemShut {NoStop}%
\bibitem [{\citenamefont {Berti}\ \emph {et~al.}(2009)\citenamefont {Berti}, \citenamefont {Cardoso},\ and\ \citenamefont {Starinets}}]{Berti_2009}%
  \BibitemOpen
  \bibfield  {author} {\bibinfo {author} {\bibfnamefont {E.}~\bibnamefont {Berti}}, \bibinfo {author} {\bibfnamefont {V.}~\bibnamefont {Cardoso}}, \ and\ \bibinfo {author} {\bibfnamefont {A.~O.}\ \bibnamefont {Starinets}},\ }\href {\doibase 10.1088/0264-9381/26/16/163001} {\bibfield  {journal} {\bibinfo  {journal} {Class. Quant. Grav.}\ }\textbf {\bibinfo {volume} {26}},\ \bibinfo {pages} {163001} (\bibinfo {year} {2009})}\BibitemShut {NoStop}%
\bibitem [{\citenamefont {Corbelli}\ and\ \citenamefont {Salucci}(2000)}]{Corbelli_2000}%
  \BibitemOpen
  \bibfield  {author} {\bibinfo {author} {\bibfnamefont {E.}~\bibnamefont {Corbelli}}\ and\ \bibinfo {author} {\bibfnamefont {P.}~\bibnamefont {Salucci}},\ }\href {\doibase 10.1046/j.1365-8711.2000.03075.x} {\bibfield  {journal} {\bibinfo  {journal} {Mon. Not. R. Astron. Soc.}\ }\textbf {\bibinfo {volume} {311}},\ \bibinfo {pages} {441} (\bibinfo {year} {2000})}\BibitemShut {NoStop}%
\bibitem [{\citenamefont {{Rogstad}}\ and\ \citenamefont {{Shostak}}(1972)}]{Rogstad1972}%
  \BibitemOpen
  \bibfield  {author} {\bibinfo {author} {\bibfnamefont {D.~H.}\ \bibnamefont {{Rogstad}}}\ and\ \bibinfo {author} {\bibfnamefont {G.~S.}\ \bibnamefont {{Shostak}}},\ }\href {\doibase 10.1086/151636} {\bibfield  {journal} {\bibinfo  {journal} {Astrophys. J.}\ }\textbf {\bibinfo {volume} {176}},\ \bibinfo {pages} {315} (\bibinfo {year} {1972})}\BibitemShut {NoStop}%
\bibitem [{\citenamefont {Weinberg}(1989)}]{RevModPhys.61.1}%
  \BibitemOpen
  \bibfield  {author} {\bibinfo {author} {\bibfnamefont {S.}~\bibnamefont {Weinberg}},\ }\href {\doibase 10.1103/RevModPhys.61.1} {\bibfield  {journal} {\bibinfo  {journal} {Rev. Mod. Phys.}\ }\textbf {\bibinfo {volume} {61}},\ \bibinfo {pages} {1} (\bibinfo {year} {1989})}\BibitemShut {NoStop}%
\bibitem [{\citenamefont {{Riess}}\ \emph {et~al.}(1998)\citenamefont {{Riess}} \emph {et~al.}}]{1998AJ....116.1009R}%
  \BibitemOpen
  \bibfield  {author} {\bibinfo {author} {\bibfnamefont {A.~G.}\ \bibnamefont {{Riess}}} \emph {et~al.},\ }\href {\doibase 10.1086/300499} {\bibfield  {journal} {\bibinfo  {journal} {Astron. J.}\ }\textbf {\bibinfo {volume} {116}},\ \bibinfo {pages} {1009} (\bibinfo {year} {1998})}\BibitemShut {NoStop}%
\bibitem [{\citenamefont {Brans}\ and\ \citenamefont {Dicke}(1961)}]{PhysRev.124.925}%
  \BibitemOpen
  \bibfield  {author} {\bibinfo {author} {\bibfnamefont {C.}~\bibnamefont {Brans}}\ and\ \bibinfo {author} {\bibfnamefont {R.~H.}\ \bibnamefont {Dicke}},\ }\href {\doibase 10.1103/PhysRev.124.925} {\bibfield  {journal} {\bibinfo  {journal} {Phys. Rev.}\ }\textbf {\bibinfo {volume} {124}},\ \bibinfo {pages} {925} (\bibinfo {year} {1961})}\BibitemShut {NoStop}%
\bibitem [{\citenamefont {Sotiriou}\ and\ \citenamefont {Faraoni}(2010)}]{RevModPhys.82.451}%
  \BibitemOpen
  \bibfield  {author} {\bibinfo {author} {\bibfnamefont {T.~P.}\ \bibnamefont {Sotiriou}}\ and\ \bibinfo {author} {\bibfnamefont {V.}~\bibnamefont {Faraoni}},\ }\href {\doibase 10.1103/revmodphys.82.451} {\bibfield  {journal} {\bibinfo  {journal} {Rev. Mod. Phys.}\ }\textbf {\bibinfo {volume} {82}},\ \bibinfo {pages} {451–497} (\bibinfo {year} {2010})}\BibitemShut {NoStop}%
\bibitem [{\citenamefont {Horndeski}(1974)}]{Horndeski:1974wa}%
  \BibitemOpen
  \bibfield  {author} {\bibinfo {author} {\bibfnamefont {G.~W.}\ \bibnamefont {Horndeski}},\ }\href {\doibase 10.1007/BF01807638} {\bibfield  {journal} {\bibinfo  {journal} {Int. J. Theor. Phys.}\ }\textbf {\bibinfo {volume} {10}},\ \bibinfo {pages} {363} (\bibinfo {year} {1974})}\BibitemShut {NoStop}%
\bibitem [{\citenamefont {Clifton}\ \emph {et~al.}(2012)\citenamefont {Clifton}, \citenamefont {Ferreira}, \citenamefont {Padilla},\ and\ \citenamefont {Skordis}}]{Clifton:2011jh}%
  \BibitemOpen
  \bibfield  {author} {\bibinfo {author} {\bibfnamefont {T.}~\bibnamefont {Clifton}}, \bibinfo {author} {\bibfnamefont {P.~G.}\ \bibnamefont {Ferreira}}, \bibinfo {author} {\bibfnamefont {A.}~\bibnamefont {Padilla}}, \ and\ \bibinfo {author} {\bibfnamefont {C.}~\bibnamefont {Skordis}},\ }\href {\doibase 10.1016/j.physrep.2012.01.001} {\bibfield  {journal} {\bibinfo  {journal} {Phys. Rept.}\ }\textbf {\bibinfo {volume} {513}},\ \bibinfo {pages} {1} (\bibinfo {year} {2012})}\BibitemShut {NoStop}%
\bibitem [{\citenamefont {Molina}\ \emph {et~al.}(2010)\citenamefont {Molina}, \citenamefont {Pani}, \citenamefont {Cardoso},\ and\ \citenamefont {Gualtieri}}]{PhysRevD.81.124021}%
  \BibitemOpen
  \bibfield  {author} {\bibinfo {author} {\bibfnamefont {C.}~\bibnamefont {Molina}}, \bibinfo {author} {\bibfnamefont {P.}~\bibnamefont {Pani}}, \bibinfo {author} {\bibfnamefont {V.}~\bibnamefont {Cardoso}}, \ and\ \bibinfo {author} {\bibfnamefont {L.}~\bibnamefont {Gualtieri}},\ }\href {\doibase 10.1103/PhysRevD.81.124021} {\bibfield  {journal} {\bibinfo  {journal} {Phys. Rev. D}\ }\textbf {\bibinfo {volume} {81}},\ \bibinfo {pages} {124021} (\bibinfo {year} {2010})}\BibitemShut {NoStop}%
\bibitem [{\citenamefont {Bl\'azquez-Salcedo}\ \emph {et~al.}(2016)\citenamefont {Bl\'azquez-Salcedo}, \citenamefont {Macedo}, \citenamefont {Cardoso}, \citenamefont {Ferrari}, \citenamefont {Gualtieri}, \citenamefont {Khoo}, \citenamefont {Kunz},\ and\ \citenamefont {Pani}}]{PhysRevD.94.104024}%
  \BibitemOpen
  \bibfield  {author} {\bibinfo {author} {\bibfnamefont {J.~L.}\ \bibnamefont {Bl\'azquez-Salcedo}}, \bibinfo {author} {\bibfnamefont {C.~F.~B.}\ \bibnamefont {Macedo}}, \bibinfo {author} {\bibfnamefont {V.}~\bibnamefont {Cardoso}}, \bibinfo {author} {\bibfnamefont {V.}~\bibnamefont {Ferrari}}, \bibinfo {author} {\bibfnamefont {L.}~\bibnamefont {Gualtieri}}, \bibinfo {author} {\bibfnamefont {F.~S.}\ \bibnamefont {Khoo}}, \bibinfo {author} {\bibfnamefont {J.}~\bibnamefont {Kunz}}, \ and\ \bibinfo {author} {\bibfnamefont {P.}~\bibnamefont {Pani}},\ }\href {\doibase 10.1103/PhysRevD.94.104024} {\bibfield  {journal} {\bibinfo  {journal} {Phys. Rev. D}\ }\textbf {\bibinfo {volume} {94}},\ \bibinfo {pages} {104024} (\bibinfo {year} {2016})}\BibitemShut {NoStop}%
\bibitem [{\citenamefont {Dreyer}\ \emph {et~al.}(2004)\citenamefont {Dreyer}, \citenamefont {Kelly}, \citenamefont {Krishnan}, \citenamefont {Finn}, \citenamefont {Garrison},\ and\ \citenamefont {Lopez-Aleman}}]{Dreyer:2003bv}%
  \BibitemOpen
  \bibfield  {author} {\bibinfo {author} {\bibfnamefont {O.}~\bibnamefont {Dreyer}}, \bibinfo {author} {\bibfnamefont {B.~J.}\ \bibnamefont {Kelly}}, \bibinfo {author} {\bibfnamefont {B.}~\bibnamefont {Krishnan}}, \bibinfo {author} {\bibfnamefont {L.~S.}\ \bibnamefont {Finn}}, \bibinfo {author} {\bibfnamefont {D.}~\bibnamefont {Garrison}}, \ and\ \bibinfo {author} {\bibfnamefont {R.}~\bibnamefont {Lopez-Aleman}},\ }\href {\doibase 10.1088/0264-9381/21/4/003} {\bibfield  {journal} {\bibinfo  {journal} {Class. Quant. Grav.}\ }\textbf {\bibinfo {volume} {21}},\ \bibinfo {pages} {787} (\bibinfo {year} {2004})}\BibitemShut {NoStop}%
\bibitem [{\citenamefont {Berti}\ \emph {et~al.}(2006)\citenamefont {Berti}, \citenamefont {Cardoso},\ and\ \citenamefont {Will}}]{berti2006gravitational}%
  \BibitemOpen
  \bibfield  {author} {\bibinfo {author} {\bibfnamefont {E.}~\bibnamefont {Berti}}, \bibinfo {author} {\bibfnamefont {V.}~\bibnamefont {Cardoso}}, \ and\ \bibinfo {author} {\bibfnamefont {C.~M.}\ \bibnamefont {Will}},\ }\href@noop {} {\bibfield  {journal} {\bibinfo  {journal} {Physical Review D}\ }\textbf {\bibinfo {volume} {73}},\ \bibinfo {pages} {064030} (\bibinfo {year} {2006})}\BibitemShut {NoStop}%
\bibitem [{\citenamefont {Lestingi}\ \emph {et~al.}(2025)\citenamefont {Lestingi}, \citenamefont {D'Addario},\ and\ \citenamefont {Sotiriou}}]{vdk4-tg61}%
  \BibitemOpen
  \bibfield  {author} {\bibinfo {author} {\bibfnamefont {J.}~\bibnamefont {Lestingi}}, \bibinfo {author} {\bibfnamefont {G.}~\bibnamefont {D'Addario}}, \ and\ \bibinfo {author} {\bibfnamefont {T.~P.}\ \bibnamefont {Sotiriou}},\ }\href {\doibase 10.1103/vdk4-tg61} {\bibfield  {journal} {\bibinfo  {journal} {Phys. Rev. D}\ }\textbf {\bibinfo {volume} {112}},\ \bibinfo {pages} {064070} (\bibinfo {year} {2025})}\BibitemShut {NoStop}%
\bibitem [{\citenamefont {Crescimbeni}\ \emph {et~al.}(2026{\natexlab{a}})\citenamefont {Crescimbeni}, \citenamefont {Forteza}, \citenamefont {Bhagwat}, \citenamefont {Westerweck},\ and\ \citenamefont {Pani}}]{SciPostPhys.20.1.025}%
  \BibitemOpen
  \bibfield  {author} {\bibinfo {author} {\bibfnamefont {F.}~\bibnamefont {Crescimbeni}}, \bibinfo {author} {\bibfnamefont {X.~J.}\ \bibnamefont {Forteza}}, \bibinfo {author} {\bibfnamefont {S.}~\bibnamefont {Bhagwat}}, \bibinfo {author} {\bibfnamefont {J.}~\bibnamefont {Westerweck}}, \ and\ \bibinfo {author} {\bibfnamefont {P.}~\bibnamefont {Pani}},\ }\href {\doibase 10.21468/SciPostPhys.20.1.025} {\bibfield  {journal} {\bibinfo  {journal} {SciPost Phys.}\ }\textbf {\bibinfo {volume} {20}},\ \bibinfo {pages} {025} (\bibinfo {year} {2026}{\natexlab{a}})}\BibitemShut {NoStop}%
\bibitem [{\citenamefont {Crescimbeni}\ \emph {et~al.}(2026{\natexlab{b}})\citenamefont {Crescimbeni}, \citenamefont {Jimenez~Forteza},\ and\ \citenamefont {Pani}}]{ptg5-f769}%
  \BibitemOpen
  \bibfield  {author} {\bibinfo {author} {\bibfnamefont {F.}~\bibnamefont {Crescimbeni}}, \bibinfo {author} {\bibfnamefont {X.}~\bibnamefont {Jimenez~Forteza}}, \ and\ \bibinfo {author} {\bibfnamefont {P.}~\bibnamefont {Pani}},\ }\href {\doibase 10.1103/ptg5-f769} {\bibfield  {journal} {\bibinfo  {journal} {Phys. Rev. D}\ }\textbf {\bibinfo {volume} {113}},\ \bibinfo {pages} {044064} (\bibinfo {year} {2026}{\natexlab{b}})}\BibitemShut {NoStop}%
\bibitem [{\citenamefont {Abac}\ \emph {et~al.}(2025{\natexlab{b}})\citenamefont {Abac} \emph {et~al.}}]{kw5g-d732}%
  \BibitemOpen
  \bibfield  {author} {\bibinfo {author} {\bibfnamefont {A.~G.}\ \bibnamefont {Abac}} \emph {et~al.} (\bibinfo {collaboration} {LIGO Scientific Collaboration, the Virgo Collaboration, and the KAGRA Collaboration}),\ }\href {\doibase 10.1103/kw5g-d732} {\bibfield  {journal} {\bibinfo  {journal} {Phys. Rev. Lett.}\ }\textbf {\bibinfo {volume} {135}},\ \bibinfo {pages} {111403} (\bibinfo {year} {2025}{\natexlab{b}})}\BibitemShut {NoStop}%
\bibitem [{\citenamefont {Abac}\ \emph {et~al.}(2026{\natexlab{c}})\citenamefont {Abac} \emph {et~al.}}]{LIGOScientific:2026wpt}%
  \BibitemOpen
  \bibfield  {author} {\bibinfo {author} {\bibfnamefont {A.~G.}\ \bibnamefont {Abac}} \emph {et~al.} (\bibinfo {collaboration} {LIGO Scientific Collaboration, the Virgo Collaboration, and the KAGRA Collaboration}),\ }\href@noop {} {\  (\bibinfo {year} {2026}{\natexlab{c}})},\ \Eprint {http://arxiv.org/abs/2603.19021} {arXiv:2603.19021 [gr-qc]} \BibitemShut {NoStop}%
\bibitem [{\citenamefont {Silva}\ \emph {et~al.}(2023)\citenamefont {Silva}, \citenamefont {Ghosh},\ and\ \citenamefont {Buonanno}}]{PhysRevD.107.044030}%
  \BibitemOpen
  \bibfield  {author} {\bibinfo {author} {\bibfnamefont {H.~O.}\ \bibnamefont {Silva}}, \bibinfo {author} {\bibfnamefont {A.}~\bibnamefont {Ghosh}}, \ and\ \bibinfo {author} {\bibfnamefont {A.}~\bibnamefont {Buonanno}},\ }\href {\doibase 10.1103/PhysRevD.107.044030} {\bibfield  {journal} {\bibinfo  {journal} {Phys. Rev. D}\ }\textbf {\bibinfo {volume} {107}},\ \bibinfo {pages} {044030} (\bibinfo {year} {2023})}\BibitemShut {NoStop}%
\bibitem [{\citenamefont {{Ka-Wai Chung}}\ and\ \citenamefont {{Yunes}}(2025)}]{2025arXiv250614695K}%
  \BibitemOpen
  \bibfield  {author} {\bibinfo {author} {\bibfnamefont {A.}~\bibnamefont {{Ka-Wai Chung}}}\ and\ \bibinfo {author} {\bibfnamefont {N.}~\bibnamefont {{Yunes}}},\ }\href {\doibase 10.48550/arXiv.2506.14695} {\bibfield  {journal} {\bibinfo  {journal} {arXiv e-prints}\ ,\ \bibinfo {eid} {arXiv:2506.14695}} (\bibinfo {year} {2025})},\ \Eprint {http://arxiv.org/abs/2506.14695} {arXiv:2506.14695 [gr-qc]} \BibitemShut {NoStop}%
\bibitem [{\citenamefont {Li}\ \emph {et~al.}(2023)\citenamefont {Li}, \citenamefont {Wagle}, \citenamefont {Chen},\ and\ \citenamefont {Yunes}}]{Li:2022pcy}%
  \BibitemOpen
  \bibfield  {author} {\bibinfo {author} {\bibfnamefont {D.}~\bibnamefont {Li}}, \bibinfo {author} {\bibfnamefont {P.}~\bibnamefont {Wagle}}, \bibinfo {author} {\bibfnamefont {Y.}~\bibnamefont {Chen}}, \ and\ \bibinfo {author} {\bibfnamefont {N.}~\bibnamefont {Yunes}},\ }\href {\doibase 10.1103/PhysRevX.13.021029} {\bibfield  {journal} {\bibinfo  {journal} {Phys. Rev. X}\ }\textbf {\bibinfo {volume} {13}},\ \bibinfo {pages} {021029} (\bibinfo {year} {2023})}\BibitemShut {NoStop}%
\bibitem [{\citenamefont {Hussain}\ and\ \citenamefont {Zimmerman}(2022)}]{Hussain:2022ins}%
  \BibitemOpen
  \bibfield  {author} {\bibinfo {author} {\bibfnamefont {A.}~\bibnamefont {Hussain}}\ and\ \bibinfo {author} {\bibfnamefont {A.}~\bibnamefont {Zimmerman}},\ }\href {\doibase 10.1103/PhysRevD.106.104018} {\bibfield  {journal} {\bibinfo  {journal} {Phys. Rev. D}\ }\textbf {\bibinfo {volume} {106}},\ \bibinfo {pages} {104018} (\bibinfo {year} {2022})}\BibitemShut {NoStop}%
\bibitem [{\citenamefont {{Teukolsky}}(1973)}]{1973ApJ...185..635T}%
  \BibitemOpen
  \bibfield  {author} {\bibinfo {author} {\bibfnamefont {S.~A.}\ \bibnamefont {{Teukolsky}}},\ }\href {\doibase 10.1086/152444} {\bibfield  {journal} {\bibinfo  {journal} {\apj}\ }\textbf {\bibinfo {volume} {185}},\ \bibinfo {pages} {635} (\bibinfo {year} {1973})}\BibitemShut {NoStop}%
\bibitem [{\citenamefont {Deffayet}\ and\ \citenamefont {Steer}(2013)}]{Deffayet_2013}%
  \BibitemOpen
  \bibfield  {author} {\bibinfo {author} {\bibfnamefont {C.}~\bibnamefont {Deffayet}}\ and\ \bibinfo {author} {\bibfnamefont {D.~A.}\ \bibnamefont {Steer}},\ }\href {\doibase 10.1088/0264-9381/30/21/214006} {\bibfield  {journal} {\bibinfo  {journal} {Class. Quantum Grav.}\ }\textbf {\bibinfo {volume} {30}},\ \bibinfo {pages} {214006} (\bibinfo {year} {2013})}\BibitemShut {NoStop}%
\bibitem [{\citenamefont {Sotiriou}\ and\ \citenamefont {Zhou}(2014{\natexlab{a}})}]{PhysRevD.90.124063}%
  \BibitemOpen
  \bibfield  {author} {\bibinfo {author} {\bibfnamefont {T.~P.}\ \bibnamefont {Sotiriou}}\ and\ \bibinfo {author} {\bibfnamefont {S.-Y.}\ \bibnamefont {Zhou}},\ }\href {\doibase 10.1103/PhysRevD.90.124063} {\bibfield  {journal} {\bibinfo  {journal} {Phys. Rev. D}\ }\textbf {\bibinfo {volume} {90}},\ \bibinfo {pages} {124063} (\bibinfo {year} {2014}{\natexlab{a}})}\BibitemShut {NoStop}%
\bibitem [{\citenamefont {Saravani}\ and\ \citenamefont {Sotiriou}(2019)}]{PhysRevD.99.124004}%
  \BibitemOpen
  \bibfield  {author} {\bibinfo {author} {\bibfnamefont {M.}~\bibnamefont {Saravani}}\ and\ \bibinfo {author} {\bibfnamefont {T.~P.}\ \bibnamefont {Sotiriou}},\ }\href {\doibase 10.1103/PhysRevD.99.124004} {\bibfield  {journal} {\bibinfo  {journal} {Phys. Rev. D}\ }\textbf {\bibinfo {volume} {99}},\ \bibinfo {pages} {124004} (\bibinfo {year} {2019})}\BibitemShut {NoStop}%
\bibitem [{\citenamefont {Sotiriou}(2015)}]{Sotiriou:2014yhm}%
  \BibitemOpen
  \bibfield  {author} {\bibinfo {author} {\bibfnamefont {T.~P.}\ \bibnamefont {Sotiriou}},\ }\href {\doibase 10.1007/978-3-319-10070-8_1} {\bibfield  {journal} {\bibinfo  {journal} {Lect. Notes Phys.}\ }\textbf {\bibinfo {volume} {892}},\ \bibinfo {pages} {3} (\bibinfo {year} {2015})}\BibitemShut {NoStop}%
\bibitem [{\citenamefont {Barack}\ \emph {et~al.}(2019)\citenamefont {Barack} \emph {et~al.}}]{Barack:2018yly}%
  \BibitemOpen
  \bibfield  {author} {\bibinfo {author} {\bibfnamefont {L.}~\bibnamefont {Barack}} \emph {et~al.},\ }\href {\doibase 10.1088/1361-6382/ab0587} {\bibfield  {journal} {\bibinfo  {journal} {Class. Quant. Grav.}\ }\textbf {\bibinfo {volume} {36}},\ \bibinfo {pages} {143001} (\bibinfo {year} {2019})}\BibitemShut {NoStop}%
\bibitem [{\citenamefont {Copeland}\ \emph {et~al.}(2006)\citenamefont {Copeland}, \citenamefont {Sami},\ and\ \citenamefont {Tsujikawa}}]{Copeland:2006wr}%
  \BibitemOpen
  \bibfield  {author} {\bibinfo {author} {\bibfnamefont {E.~J.}\ \bibnamefont {Copeland}}, \bibinfo {author} {\bibfnamefont {M.}~\bibnamefont {Sami}}, \ and\ \bibinfo {author} {\bibfnamefont {S.}~\bibnamefont {Tsujikawa}},\ }\href {\doibase 10.1142/S021827180600942X} {\bibfield  {journal} {\bibinfo  {journal} {Int. J. Mod. Phys. D}\ }\textbf {\bibinfo {volume} {15}},\ \bibinfo {pages} {1753} (\bibinfo {year} {2006})}\BibitemShut {NoStop}%
\bibitem [{\citenamefont {Kim}\ and\ \citenamefont {Carosi}(2010)}]{Kim:2008hd}%
  \BibitemOpen
  \bibfield  {author} {\bibinfo {author} {\bibfnamefont {J.~E.}\ \bibnamefont {Kim}}\ and\ \bibinfo {author} {\bibfnamefont {G.}~\bibnamefont {Carosi}},\ }\href {\doibase 10.1103/RevModPhys.82.557} {\bibfield  {journal} {\bibinfo  {journal} {Rev. Mod. Phys.}\ }\textbf {\bibinfo {volume} {82}},\ \bibinfo {pages} {557} (\bibinfo {year} {2010})},\ \bibinfo {note} {[Erratum: Rev.Mod.Phys. 91, 049902 (2019)]}\BibitemShut {NoStop}%
\bibitem [{\citenamefont {Marsh}(2016)}]{Marsh:2015xka}%
  \BibitemOpen
  \bibfield  {author} {\bibinfo {author} {\bibfnamefont {D.~J.~E.}\ \bibnamefont {Marsh}},\ }\href {\doibase 10.1016/j.physrep.2016.06.005} {\bibfield  {journal} {\bibinfo  {journal} {Phys. Rept.}\ }\textbf {\bibinfo {volume} {643}},\ \bibinfo {pages} {1} (\bibinfo {year} {2016})}\BibitemShut {NoStop}%
\bibitem [{\citenamefont {Hui}(2021)}]{Hui:2021tkt}%
  \BibitemOpen
  \bibfield  {author} {\bibinfo {author} {\bibfnamefont {L.}~\bibnamefont {Hui}},\ }\href {\doibase 10.1146/annurev-astro-120920-010024} {\bibfield  {journal} {\bibinfo  {journal} {Ann. Rev. Astron. Astrophys.}\ }\textbf {\bibinfo {volume} {59}},\ \bibinfo {pages} {247} (\bibinfo {year} {2021})}\BibitemShut {NoStop}%
\bibitem [{\citenamefont {D'Addario}\ \emph {et~al.}(2024)\citenamefont {D'Addario}, \citenamefont {Padilla}, \citenamefont {Saffin}, \citenamefont {Sotiriou},\ and\ \citenamefont {Spiers}}]{PhysRevD.109.084046}%
  \BibitemOpen
  \bibfield  {author} {\bibinfo {author} {\bibfnamefont {G.}~\bibnamefont {D'Addario}}, \bibinfo {author} {\bibfnamefont {A.}~\bibnamefont {Padilla}}, \bibinfo {author} {\bibfnamefont {P.~M.}\ \bibnamefont {Saffin}}, \bibinfo {author} {\bibfnamefont {T.~P.}\ \bibnamefont {Sotiriou}}, \ and\ \bibinfo {author} {\bibfnamefont {A.}~\bibnamefont {Spiers}},\ }\href {\doibase 10.1103/PhysRevD.109.084046} {\bibfield  {journal} {\bibinfo  {journal} {Phys. Rev. D}\ }\textbf {\bibinfo {volume} {109}},\ \bibinfo {pages} {084046} (\bibinfo {year} {2024})}\BibitemShut {NoStop}%
\bibitem [{\citenamefont {Hawking}(1972)}]{Hawking:1972qk}%
  \BibitemOpen
  \bibfield  {author} {\bibinfo {author} {\bibfnamefont {S.~W.}\ \bibnamefont {Hawking}},\ }\href {\doibase 10.1007/BF01877518} {\bibfield  {journal} {\bibinfo  {journal} {Commun. Math. Phys.}\ }\textbf {\bibinfo {volume} {25}},\ \bibinfo {pages} {167} (\bibinfo {year} {1972})}\BibitemShut {NoStop}%
\bibitem [{\citenamefont {Sotiriou}\ and\ \citenamefont {Faraoni}(2012)}]{PhysRevLett.108.081103}%
  \BibitemOpen
  \bibfield  {author} {\bibinfo {author} {\bibfnamefont {T.~P.}\ \bibnamefont {Sotiriou}}\ and\ \bibinfo {author} {\bibfnamefont {V.}~\bibnamefont {Faraoni}},\ }\href {\doibase 10.1103/PhysRevLett.108.081103} {\bibfield  {journal} {\bibinfo  {journal} {Phys. Rev. Lett.}\ }\textbf {\bibinfo {volume} {108}},\ \bibinfo {pages} {081103} (\bibinfo {year} {2012})}\BibitemShut {NoStop}%
\bibitem [{\citenamefont {Herdeiro}\ and\ \citenamefont {Radu}(2015)}]{doi:10.1142/S0218271815420146}%
  \BibitemOpen
  \bibfield  {author} {\bibinfo {author} {\bibfnamefont {C.~A.~R.}\ \bibnamefont {Herdeiro}}\ and\ \bibinfo {author} {\bibfnamefont {E.}~\bibnamefont {Radu}},\ }\href {\doibase 10.1142/S0218271815420146} {\bibfield  {journal} {\bibinfo  {journal} {Int. J. Mod. Phys. D}\ }\textbf {\bibinfo {volume} {24}},\ \bibinfo {pages} {1542014} (\bibinfo {year} {2015})}\BibitemShut {NoStop}%
\bibitem [{\citenamefont {Hui}\ and\ \citenamefont {Nicolis}(2013)}]{Hui:2012qt}%
  \BibitemOpen
  \bibfield  {author} {\bibinfo {author} {\bibfnamefont {L.}~\bibnamefont {Hui}}\ and\ \bibinfo {author} {\bibfnamefont {A.}~\bibnamefont {Nicolis}},\ }\href {\doibase 10.1103/PhysRevLett.110.241104} {\bibfield  {journal} {\bibinfo  {journal} {Phys. Rev. Lett.}\ }\textbf {\bibinfo {volume} {110}},\ \bibinfo {pages} {241104} (\bibinfo {year} {2013})}\BibitemShut {NoStop}%
\bibitem [{\citenamefont {Sotiriou}\ and\ \citenamefont {Zhou}(2014{\natexlab{b}})}]{PhysRevLett.112.251102}%
  \BibitemOpen
  \bibfield  {author} {\bibinfo {author} {\bibfnamefont {T.~P.}\ \bibnamefont {Sotiriou}}\ and\ \bibinfo {author} {\bibfnamefont {S.-Y.}\ \bibnamefont {Zhou}},\ }\href {\doibase 10.1103/PhysRevLett.112.251102} {\bibfield  {journal} {\bibinfo  {journal} {Phys. Rev. Lett.}\ }\textbf {\bibinfo {volume} {112}},\ \bibinfo {pages} {251102} (\bibinfo {year} {2014}{\natexlab{b}})}\BibitemShut {NoStop}%
\bibitem [{\citenamefont {Thaalba}\ \emph {et~al.}(2023)\citenamefont {Thaalba}, \citenamefont {Antoniou},\ and\ \citenamefont {Sotiriou}}]{Thaalba:2022bnt}%
  \BibitemOpen
  \bibfield  {author} {\bibinfo {author} {\bibfnamefont {F.}~\bibnamefont {Thaalba}}, \bibinfo {author} {\bibfnamefont {G.}~\bibnamefont {Antoniou}}, \ and\ \bibinfo {author} {\bibfnamefont {T.~P.}\ \bibnamefont {Sotiriou}},\ }\href {\doibase 10.1088/1361-6382/acdd42} {\bibfield  {journal} {\bibinfo  {journal} {Class. Quant. Grav.}\ }\textbf {\bibinfo {volume} {40}},\ \bibinfo {pages} {155002} (\bibinfo {year} {2023})}\BibitemShut {NoStop}%
\bibitem [{\citenamefont {Kobayashi}\ \emph {et~al.}(2011)\citenamefont {Kobayashi}, \citenamefont {Yamaguchi},\ and\ \citenamefont {Yokoyama}}]{10.1143/PTP.126.511}%
  \BibitemOpen
  \bibfield  {author} {\bibinfo {author} {\bibfnamefont {T.}~\bibnamefont {Kobayashi}}, \bibinfo {author} {\bibfnamefont {M.}~\bibnamefont {Yamaguchi}}, \ and\ \bibinfo {author} {\bibfnamefont {J.}~\bibnamefont {Yokoyama}},\ }\href {\doibase 10.1143/PTP.126.511} {\bibfield  {journal} {\bibinfo  {journal} {Prog. Theor. Phys.}\ }\textbf {\bibinfo {volume} {126}},\ \bibinfo {pages} {511} (\bibinfo {year} {2011})}\BibitemShut {NoStop}%
\bibitem [{\citenamefont {Tanahashi}\ and\ \citenamefont {Ohashi}(2017)}]{Tanahashi_2017}%
  \BibitemOpen
  \bibfield  {author} {\bibinfo {author} {\bibfnamefont {N.}~\bibnamefont {Tanahashi}}\ and\ \bibinfo {author} {\bibfnamefont {S.}~\bibnamefont {Ohashi}},\ }\href {\doibase 10.1088/1361-6382/aa85fb} {\bibfield  {journal} {\bibinfo  {journal} {Class. Quantum Grav.}\ }\textbf {\bibinfo {volume} {34}},\ \bibinfo {pages} {215003} (\bibinfo {year} {2017})}\BibitemShut {NoStop}%
\bibitem [{\citenamefont {{Thaalba}}\ \emph {et~al.}(2025)\citenamefont {{Thaalba}}, \citenamefont {{Gualtieri}}, \citenamefont {{Sotiriou}},\ and\ \citenamefont {{Trincherini}}}]{thaalba2025screeningdipolaremissiontwoscale}%
  \BibitemOpen
  \bibfield  {author} {\bibinfo {author} {\bibfnamefont {F.}~\bibnamefont {{Thaalba}}}, \bibinfo {author} {\bibfnamefont {L.}~\bibnamefont {{Gualtieri}}}, \bibinfo {author} {\bibfnamefont {T.~P.}\ \bibnamefont {{Sotiriou}}}, \ and\ \bibinfo {author} {\bibfnamefont {E.}~\bibnamefont {{Trincherini}}},\ }\href {\doibase 10.48550/arXiv.2512.04083} {\bibfield  {journal} {\bibinfo  {journal} {arXiv e-prints}\ ,\ \bibinfo {pages} {arXiv:2512.04083}} (\bibinfo {year} {2025})}\BibitemShut {NoStop}%
\bibitem [{\citenamefont {Spiers}\ \emph {et~al.}(2023)\citenamefont {Spiers}, \citenamefont {Pound},\ and\ \citenamefont {Moxon}}]{PhysRevD.108.064002}%
  \BibitemOpen
  \bibfield  {author} {\bibinfo {author} {\bibfnamefont {A.}~\bibnamefont {Spiers}}, \bibinfo {author} {\bibfnamefont {A.}~\bibnamefont {Pound}}, \ and\ \bibinfo {author} {\bibfnamefont {J.}~\bibnamefont {Moxon}},\ }\href {\doibase 10.1103/PhysRevD.108.064002} {\bibfield  {journal} {\bibinfo  {journal} {Phys. Rev. D}\ }\textbf {\bibinfo {volume} {108}},\ \bibinfo {pages} {064002} (\bibinfo {year} {2023})}\BibitemShut {NoStop}%
\bibitem [{\citenamefont {Spiers}\ \emph {et~al.}(2024)\citenamefont {Spiers}, \citenamefont {Maselli},\ and\ \citenamefont {Sotiriou}}]{PhysRevD.109.064022}%
  \BibitemOpen
  \bibfield  {author} {\bibinfo {author} {\bibfnamefont {A.}~\bibnamefont {Spiers}}, \bibinfo {author} {\bibfnamefont {A.}~\bibnamefont {Maselli}}, \ and\ \bibinfo {author} {\bibfnamefont {T.~P.}\ \bibnamefont {Sotiriou}},\ }\href {\doibase 10.1103/PhysRevD.109.064022} {\bibfield  {journal} {\bibinfo  {journal} {Phys. Rev. D}\ }\textbf {\bibinfo {volume} {109}},\ \bibinfo {pages} {064022} (\bibinfo {year} {2024})}\BibitemShut {NoStop}%
\bibitem [{\citenamefont {Li}\ \emph {et~al.}(2024)\citenamefont {Li}, \citenamefont {Hussain}, \citenamefont {Wagle}, \citenamefont {Chen}, \citenamefont {Yunes},\ and\ \citenamefont {Zimmerman}}]{Li:2023ulk}%
  \BibitemOpen
  \bibfield  {author} {\bibinfo {author} {\bibfnamefont {D.}~\bibnamefont {Li}}, \bibinfo {author} {\bibfnamefont {A.}~\bibnamefont {Hussain}}, \bibinfo {author} {\bibfnamefont {P.}~\bibnamefont {Wagle}}, \bibinfo {author} {\bibfnamefont {Y.}~\bibnamefont {Chen}}, \bibinfo {author} {\bibfnamefont {N.}~\bibnamefont {Yunes}}, \ and\ \bibinfo {author} {\bibfnamefont {A.}~\bibnamefont {Zimmerman}},\ }\href {\doibase 10.1103/PhysRevD.109.104026} {\bibfield  {journal} {\bibinfo  {journal} {Phys. Rev. D}\ }\textbf {\bibinfo {volume} {109}},\ \bibinfo {pages} {104026} (\bibinfo {year} {2024})}\BibitemShut {NoStop}%
\bibitem [{\citenamefont {Juli{\'e}}\ \emph {et~al.}(2025)\citenamefont {Juli{\'e}}, \citenamefont {Pompili},\ and\ \citenamefont {Buonanno}}]{Julie:2024fwy}%
  \BibitemOpen
  \bibfield  {author} {\bibinfo {author} {\bibfnamefont {F.-L.}\ \bibnamefont {Juli{\'e}}}, \bibinfo {author} {\bibfnamefont {L.}~\bibnamefont {Pompili}}, \ and\ \bibinfo {author} {\bibfnamefont {A.}~\bibnamefont {Buonanno}},\ }\href {\doibase 10.1103/PhysRevD.111.024016} {\bibfield  {journal} {\bibinfo  {journal} {Phys. Rev. D}\ }\textbf {\bibinfo {volume} {111}},\ \bibinfo {pages} {024016} (\bibinfo {year} {2025})}\BibitemShut {NoStop}%
\bibitem [{\citenamefont {Gao}\ \emph {et~al.}(2024)\citenamefont {Gao}, \citenamefont {Tang}, \citenamefont {Wang}, \citenamefont {Yan},\ and\ \citenamefont {Fan}}]{PhysRevD.110.044022}%
  \BibitemOpen
  \bibfield  {author} {\bibinfo {author} {\bibfnamefont {B.}~\bibnamefont {Gao}}, \bibinfo {author} {\bibfnamefont {S.-P.}\ \bibnamefont {Tang}}, \bibinfo {author} {\bibfnamefont {H.-T.}\ \bibnamefont {Wang}}, \bibinfo {author} {\bibfnamefont {J.}~\bibnamefont {Yan}}, \ and\ \bibinfo {author} {\bibfnamefont {Y.-Z.}\ \bibnamefont {Fan}},\ }\href {\doibase 10.1103/PhysRevD.110.044022} {\bibfield  {journal} {\bibinfo  {journal} {Phys. Rev. D}\ }\textbf {\bibinfo {volume} {110}},\ \bibinfo {pages} {044022} (\bibinfo {year} {2024})}\BibitemShut {NoStop}%
\bibitem [{\citenamefont {S\"anger}\ \emph {et~al.}(2026)\citenamefont {S\"anger} \emph {et~al.}}]{r43k-51yq}%
  \BibitemOpen
  \bibfield  {author} {\bibinfo {author} {\bibfnamefont {E.}~\bibnamefont {S\"anger}} \emph {et~al.},\ }\href {\doibase 10.1103/r43k-51yq} {\bibfield  {journal} {\bibinfo  {journal} {Phys. Rev. D}\ }\textbf {\bibinfo {volume} {113}},\ \bibinfo {pages} {084070} (\bibinfo {year} {2026})}\BibitemShut {NoStop}%
\bibitem [{\citenamefont {Maselli}\ \emph {et~al.}(2020)\citenamefont {Maselli}, \citenamefont {Franchini}, \citenamefont {Gualtieri},\ and\ \citenamefont {Sotiriou}}]{Maselli:2020zgv}%
  \BibitemOpen
  \bibfield  {author} {\bibinfo {author} {\bibfnamefont {A.}~\bibnamefont {Maselli}}, \bibinfo {author} {\bibfnamefont {N.}~\bibnamefont {Franchini}}, \bibinfo {author} {\bibfnamefont {L.}~\bibnamefont {Gualtieri}}, \ and\ \bibinfo {author} {\bibfnamefont {T.~P.}\ \bibnamefont {Sotiriou}},\ }\href {\doibase 10.1103/PhysRevLett.125.141101} {\bibfield  {journal} {\bibinfo  {journal} {Phys. Rev. Lett.}\ }\textbf {\bibinfo {volume} {125}},\ \bibinfo {pages} {141101} (\bibinfo {year} {2020})}\BibitemShut {NoStop}%
\bibitem [{\citenamefont {Maselli}\ \emph {et~al.}(2022)\citenamefont {Maselli}, \citenamefont {Franchini}, \citenamefont {Gualtieri}, \citenamefont {Sotiriou}, \citenamefont {Barsanti},\ and\ \citenamefont {Pani}}]{Maselli:2021men}%
  \BibitemOpen
  \bibfield  {author} {\bibinfo {author} {\bibfnamefont {A.}~\bibnamefont {Maselli}}, \bibinfo {author} {\bibfnamefont {N.}~\bibnamefont {Franchini}}, \bibinfo {author} {\bibfnamefont {L.}~\bibnamefont {Gualtieri}}, \bibinfo {author} {\bibfnamefont {T.~P.}\ \bibnamefont {Sotiriou}}, \bibinfo {author} {\bibfnamefont {S.}~\bibnamefont {Barsanti}}, \ and\ \bibinfo {author} {\bibfnamefont {P.}~\bibnamefont {Pani}},\ }\href {\doibase 10.1038/s41550-021-01589-5} {\bibfield  {journal} {\bibinfo  {journal} {Nature Astron.}\ }\textbf {\bibinfo {volume} {6}},\ \bibinfo {pages} {464} (\bibinfo {year} {2022})}\BibitemShut {NoStop}%
\bibitem [{\citenamefont {Speri}\ \emph {et~al.}(2026)\citenamefont {Speri}, \citenamefont {Barsanti}, \citenamefont {Maselli}, \citenamefont {Sotiriou}, \citenamefont {Warburton}, \citenamefont {van~de Meent}, \citenamefont {Chua}, \citenamefont {Burke},\ and\ \citenamefont {Gair}}]{Speri:2024qak}%
  \BibitemOpen
  \bibfield  {author} {\bibinfo {author} {\bibfnamefont {L.}~\bibnamefont {Speri}}, \bibinfo {author} {\bibfnamefont {S.}~\bibnamefont {Barsanti}}, \bibinfo {author} {\bibfnamefont {A.}~\bibnamefont {Maselli}}, \bibinfo {author} {\bibfnamefont {T.~P.}\ \bibnamefont {Sotiriou}}, \bibinfo {author} {\bibfnamefont {N.}~\bibnamefont {Warburton}}, \bibinfo {author} {\bibfnamefont {M.}~\bibnamefont {van~de Meent}}, \bibinfo {author} {\bibfnamefont {A.~J.~K.}\ \bibnamefont {Chua}}, \bibinfo {author} {\bibfnamefont {O.}~\bibnamefont {Burke}}, \ and\ \bibinfo {author} {\bibfnamefont {J.}~\bibnamefont {Gair}},\ }\href {\doibase 10.1103/cnhz-6zlk} {\bibfield  {journal} {\bibinfo  {journal} {Phys. Rev. D}\ }\textbf {\bibinfo {volume} {113}},\ \bibinfo {pages} {023036} (\bibinfo {year} {2026})}\BibitemShut {NoStop}%
\bibitem [{\citenamefont {Silva}\ \emph {et~al.}(2018)\citenamefont {Silva}, \citenamefont {Sakstein}, \citenamefont {Gualtieri}, \citenamefont {Sotiriou},\ and\ \citenamefont {Berti}}]{Silva:2017uqg}%
  \BibitemOpen
  \bibfield  {author} {\bibinfo {author} {\bibfnamefont {H.~O.}\ \bibnamefont {Silva}}, \bibinfo {author} {\bibfnamefont {J.}~\bibnamefont {Sakstein}}, \bibinfo {author} {\bibfnamefont {L.}~\bibnamefont {Gualtieri}}, \bibinfo {author} {\bibfnamefont {T.~P.}\ \bibnamefont {Sotiriou}}, \ and\ \bibinfo {author} {\bibfnamefont {E.}~\bibnamefont {Berti}},\ }\href {\doibase 10.1103/PhysRevLett.120.131104} {\bibfield  {journal} {\bibinfo  {journal} {Phys. Rev. Lett.}\ }\textbf {\bibinfo {volume} {120}},\ \bibinfo {pages} {131104} (\bibinfo {year} {2018})}\BibitemShut {NoStop}%
\bibitem [{\citenamefont {Doneva}\ and\ \citenamefont {Yazadjiev}(2018)}]{Doneva:2017bvd}%
  \BibitemOpen
  \bibfield  {author} {\bibinfo {author} {\bibfnamefont {D.~D.}\ \bibnamefont {Doneva}}\ and\ \bibinfo {author} {\bibfnamefont {S.~S.}\ \bibnamefont {Yazadjiev}},\ }\href {\doibase 10.1103/PhysRevLett.120.131103} {\bibfield  {journal} {\bibinfo  {journal} {Phys. Rev. Lett.}\ }\textbf {\bibinfo {volume} {120}},\ \bibinfo {pages} {131103} (\bibinfo {year} {2018})}\BibitemShut {NoStop}%
\bibitem [{\citenamefont {Ventagli}\ \emph {et~al.}(2020)\citenamefont {Ventagli}, \citenamefont {Leh{\'e}bel},\ and\ \citenamefont {Sotiriou}}]{Ventagli:2020rnx}%
  \BibitemOpen
  \bibfield  {author} {\bibinfo {author} {\bibfnamefont {G.}~\bibnamefont {Ventagli}}, \bibinfo {author} {\bibfnamefont {A.}~\bibnamefont {Leh{\'e}bel}}, \ and\ \bibinfo {author} {\bibfnamefont {T.~P.}\ \bibnamefont {Sotiriou}},\ }\href {\doibase 10.1103/PhysRevD.102.024050} {\bibfield  {journal} {\bibinfo  {journal} {Phys. Rev. D}\ }\textbf {\bibinfo {volume} {102}},\ \bibinfo {pages} {024050} (\bibinfo {year} {2020})}\BibitemShut {NoStop}%
\bibitem [{\citenamefont {Dima}\ \emph {et~al.}(2020)\citenamefont {Dima}, \citenamefont {Barausse}, \citenamefont {Franchini},\ and\ \citenamefont {Sotiriou}}]{PhysRevLett.125.231101}%
  \BibitemOpen
  \bibfield  {author} {\bibinfo {author} {\bibfnamefont {A.}~\bibnamefont {Dima}}, \bibinfo {author} {\bibfnamefont {E.}~\bibnamefont {Barausse}}, \bibinfo {author} {\bibfnamefont {N.}~\bibnamefont {Franchini}}, \ and\ \bibinfo {author} {\bibfnamefont {T.~P.}\ \bibnamefont {Sotiriou}},\ }\href {\doibase 10.1103/PhysRevLett.125.231101} {\bibfield  {journal} {\bibinfo  {journal} {Phys. Rev. Lett.}\ }\textbf {\bibinfo {volume} {125}},\ \bibinfo {pages} {231101} (\bibinfo {year} {2020})}\BibitemShut {NoStop}%
\bibitem [{\citenamefont {Herdeiro}\ \emph {et~al.}(2021)\citenamefont {Herdeiro}, \citenamefont {Radu}, \citenamefont {Silva}, \citenamefont {Sotiriou},\ and\ \citenamefont {Yunes}}]{PhysRevLett.126.011103}%
  \BibitemOpen
  \bibfield  {author} {\bibinfo {author} {\bibfnamefont {C.~A.~R.}\ \bibnamefont {Herdeiro}}, \bibinfo {author} {\bibfnamefont {E.}~\bibnamefont {Radu}}, \bibinfo {author} {\bibfnamefont {H.~O.}\ \bibnamefont {Silva}}, \bibinfo {author} {\bibfnamefont {T.~P.}\ \bibnamefont {Sotiriou}}, \ and\ \bibinfo {author} {\bibfnamefont {N.}~\bibnamefont {Yunes}},\ }\href {\doibase 10.1103/PhysRevLett.126.011103} {\bibfield  {journal} {\bibinfo  {journal} {Phys. Rev. Lett.}\ }\textbf {\bibinfo {volume} {126}},\ \bibinfo {pages} {011103} (\bibinfo {year} {2021})}\BibitemShut {NoStop}%
\bibitem [{\citenamefont {Doneva}\ \emph {et~al.}(2024)\citenamefont {Doneva}, \citenamefont {Ramazanoğlu}, \citenamefont {Silva}, \citenamefont {Sotiriou},\ and\ \citenamefont {Yazadjiev}}]{RevModPhys.96.015004}%
  \BibitemOpen
  \bibfield  {author} {\bibinfo {author} {\bibfnamefont {D.~D.}\ \bibnamefont {Doneva}}, \bibinfo {author} {\bibfnamefont {F.~M.}\ \bibnamefont {Ramazanoğlu}}, \bibinfo {author} {\bibfnamefont {H.~O.}\ \bibnamefont {Silva}}, \bibinfo {author} {\bibfnamefont {T.~P.}\ \bibnamefont {Sotiriou}}, \ and\ \bibinfo {author} {\bibfnamefont {S.~S.}\ \bibnamefont {Yazadjiev}},\ }\href {http://dx.doi.org/10.1103/RevModPhys.96.015004} {\bibfield  {journal} {\bibinfo  {journal} {Rev. Mod. Phys.}\ }\textbf {\bibinfo {volume} {96}} (\bibinfo {year} {2024})}\BibitemShut {NoStop}%
\end{thebibliography}%

\end{document}